\documentclass[twocolumn]{aastex61} 
\pdfoutput=1 

\usepackage{color}
\usepackage{amsmath}
\usepackage{mathtools}
\usepackage{hyperref}
\usepackage[T1]{fontenc}
\usepackage{apjfonts} 
\usepackage[figure,figure*]{hypcap}

\newcommand{\voigtshear}{\delta{}M}
\newcommand{\maxshear}{M_{U}}
\newcommand{\maxcomp}{J_{U}}
\newcommand{\maxvisc}{\eta_{S}}
\newcommand{\voigtvisc}{\eta_{P}}
\newcommand{\voigtcomp}{\delta{}J}
\newcommand{\andvisc}{\eta_{S}}
\newcommand{\andcomp}{J_{U}}

\usepackage{rotating}

\shorttitle{Tidal Dissipation using Advanced Rheologies} 
\shortauthors{J. P. Renaud and W. G. Henning}

\begin{document}

\title{Increased Tidal Dissipation using Advanced Rheological Models: Implications for Io and Tidally Active Exoplanets}

\author{\href{mailto:joe.p.renaud@gmail.com}{Joe P. Renaud}}
\affiliation{Department of Physics and Astronomy, George Mason University, Fairfax, VA, USA}
\email{jrenaud@gmu.edu}
\author{\href{mailto:wade.g.henning@nasa.gov}{Wade G. Henning}}
\affiliation{NASA Goddard Space Flight Center, Greenbelt, MD, USA}
\affiliation{Department of Astronomy, University of Maryland, College Park, MD}


\begin{abstract} 
The advanced rheological models of \citet{Andrade1910} and \citet{SundbergCooper2010} are compared to the traditional Maxwell model to understand how each affects the tidal dissipation of heat within rocky bodies. We find both the Andrade and Sundberg--Cooper rheologies can produce at least 10$\times$ the tidal heating compared to a traditional Maxwell model for a warm (1400--1600 K) Io-like satellite. Sundberg--Cooper can cause even larger dissipation around a critical temperature and frequency. These models allow cooler planets to stay tidally active in the face of orbital perturbations---a condition we term `tidal resilience.' This has implications for the time evolution of tidally active worlds, and the long-term equilibria they fall into. For instance, if Io's interior is better modeled by the Andrade or Sundberg--Cooper rheologies, the number of possible resonance-forming scenarios that still produce a hot, modern Io is expanded, and these scenarios do not require an early formation of the Laplace resonance. The two primary empirical parameters that define the Andrade anelasticity are examined in several phase spaces to provide guidance on how their uncertainties impact tidal outcomes, as laboratory studies continue to constrain their real values. We provide detailed reference tables on the fully general equations required for others to insert the Andrade and Sundberg--Cooper models into standard tidal formulae. Lastly, we show that advanced rheologies greatly impact the heating of short-period exoplanets and exomoons, while the properties of tidal resilience can mean a greater number of tidally active worlds among all extrasolar systems.
\end{abstract}


\section{Introduction}\label{sec:intro}
The way in which a planetary body responds to any non-negligible tidal forces can greatly impact its orbital and thermal evolution. It is well known that certain orbital configurations lead to large, long-lasting, tidal stresses within solar system bodies \citep[e.g.,][]{Peale1979, Cassen1980}. Indeed, some of these bodies exhibit such large stress variations that the resultant heat generation is easily detected \citep{Morabito1979}. Understanding such tidal evolution provides insights into a planet's past and future orbit, and may have implications for astrobiology.\par

In the past, the field of planetary tidal dynamics has been moderately decoupled from the nuances of laboratory material science. New work \citep[e.g.,][]{Tobie2008, Henning2009, Castillo-Rogez2012, BehounkovaCadek2014, Correia2014, HenningHurford2014, Kuchta2015, Frouard2016} has attempted to better marry the two fields through rigorous modeling of planetary geometry and composition. Recent work into the study of a planet's bulk response to stresses, or rheology, focuses on empirical models developed around laboratory studies of rock that still retain a basis in microphysical processes. Since tidal stresses in satellite bodies are expected to occur at frequencies too low for a purely elastic response, and too fast to be dominated by steady-state viscous creep, then any response model needs to accurately describe the transition between the two. This transient creep is described by both recoverable (anelastic) and non-recoverable (viscoelastic) ductile motion of a planet's bulk. The majority of prior tidal analyses have focused on rheological models such as the constant-response approach, or the Maxwell rheology. The Maxwell rheology includes only an elastic and steady-state creep response, with no transient creep regime. A first stage in improvement may be obtained by considering the Burgers rheology, which includes transient creep, but has historically had difficulty in matching Earth observations that probe the interior, such as investigations of postglacial rebound. Greater success has been obtained from the Andrade rheology \citep{Andrade1910, Jackson1993}, in part because it is founded upon laboratory experiments. For this reason, a growing body of work has now applied the Andrade rheology to planetary tidal problems including Iapetus \citep[e.g.,][]{Castillo-Rogez2011}, exoplanets \citep{ShojiKurita2014}, and Io \citep{BiersonNimmo2016}. However, to the authors' knowledge, there has not been a comprehensive comparison made between traditional models and Andrade in all applicable phase spaces.\par

As we shall show, the differences between models can be dramatic---knowing when one model is more appropriate will be critical for future planetary studies. Models beyond Andrade exist, and in this work we explore the behavior of a uniquely valuable composite model described in detail by \citet{SundbergCooper2010}, which we refer to as the Sundberg--Cooper rheology. The experimental success that the Andrade rheology, or its cousin Sundburg-Cooper, has had in describing grain boundary processes is very promising for modeling transient creep in both rock and ice \citep[e.g.,][]{SundbergCooper2010, FaulJackson2015b, McCarthyCooper2016}.\par

We present an analysis of a large phase space relevant to planetary tidal physics to better constrain when transient rheologies differ significantly from the traditionally used non-transient Maxwell model. We also examine the impact that attenuation flattening, exhibited by the Andrade and Sundberg--Cooper models, has compared to the specific peaks found in a Burgers-like model. First, this analysis is conducted on a hypothetical system that is subjected to tidal stresses. To give this system context we set many of the parameters to mimic the Io--Jupiter system (see Section \ref{sec:methods}). We find that the transient response exhibited by the Andrade mechanism greatly influences low-temperature and/or high-frequency dissipation. Secular cooling drives mantles into this high-dissipation region, thereby impacting a planet's thermal evolution and possible equilibrium. We also present comprehensive tables of the relevant governing equations, many newly derived in this work, as a reference resource. In Section \ref{sec:results:exoplanets}, we extend the analysis from Io to parameter ranges encompassing observed terrestrial-class extrasolar planets, to demonstrate how the enhancements of tidal activity by the Andrade and Sundberg--Cooper models will alter such objects.\par

\section{Background} \label{sec:background}
A rich history of tidal investigation has provided the foundation for the work outlined here \citep[e.g.,][]{Darwin1880,Kaula1964,GoldreichSoter1966, Hut1981, Ferraz-Mello2008, EfroimskyMakarov2014}. Tidal forces are generated by a non-zero gravitational potential gradient throughout a satellite. These forces lead to internal stress, which is counteracted by the satellite's material strength. Variation of this gradient in time, due to either an eccentric orbit, a non-synchronous rotation (NSR), a non-zero obliquity, or some combination, leads to frictional dissipation of orbital and/or spin energy into internal heat. \par

Spin--orbit resonances, and resonances with other satellites' orbits, can pump a satellite's eccentricity or force an NSR state. These bodies will then experience an exchange of some of this pumped orbital/spin energy into heat via tidal interactions \citep[][]{MD}. The continuous pumping can lead to extended periods of significant tidal dissipation, such as that seen on Io \citep[e.g.,][]{HussmannSpohn2004}. \par

In this study we do not explicitly consider tidal heating in fluid layers \citep{Tyler2008, Tyler2009, Matsuyama2014}. Such heating may play a central role for Io \citep{TylerHenningHamilton2015}, if a conducting subsurface magma slush layer exists \citep{Khurana2011}. However, even if fluid heating is ongoing, its contribution sums linearly with solid-body tides, meaning that all issues raised in this report remain equally valid. In particular, the majority of effects we discuss have to do with cold-end-member Io conditions such as may occur in low-eccentricity excursions, or before the onset of the Laplace resonance (see Section \ref{sec:results:laplace}). In these situations a magma ocean would not even exist, and solid-body tides become even more important.

\subsection{Material Physics}\label{sec:mat_physics}

\begin{figure*}[hbt!]
    \centering
    \includegraphics[width=0.8\linewidth]{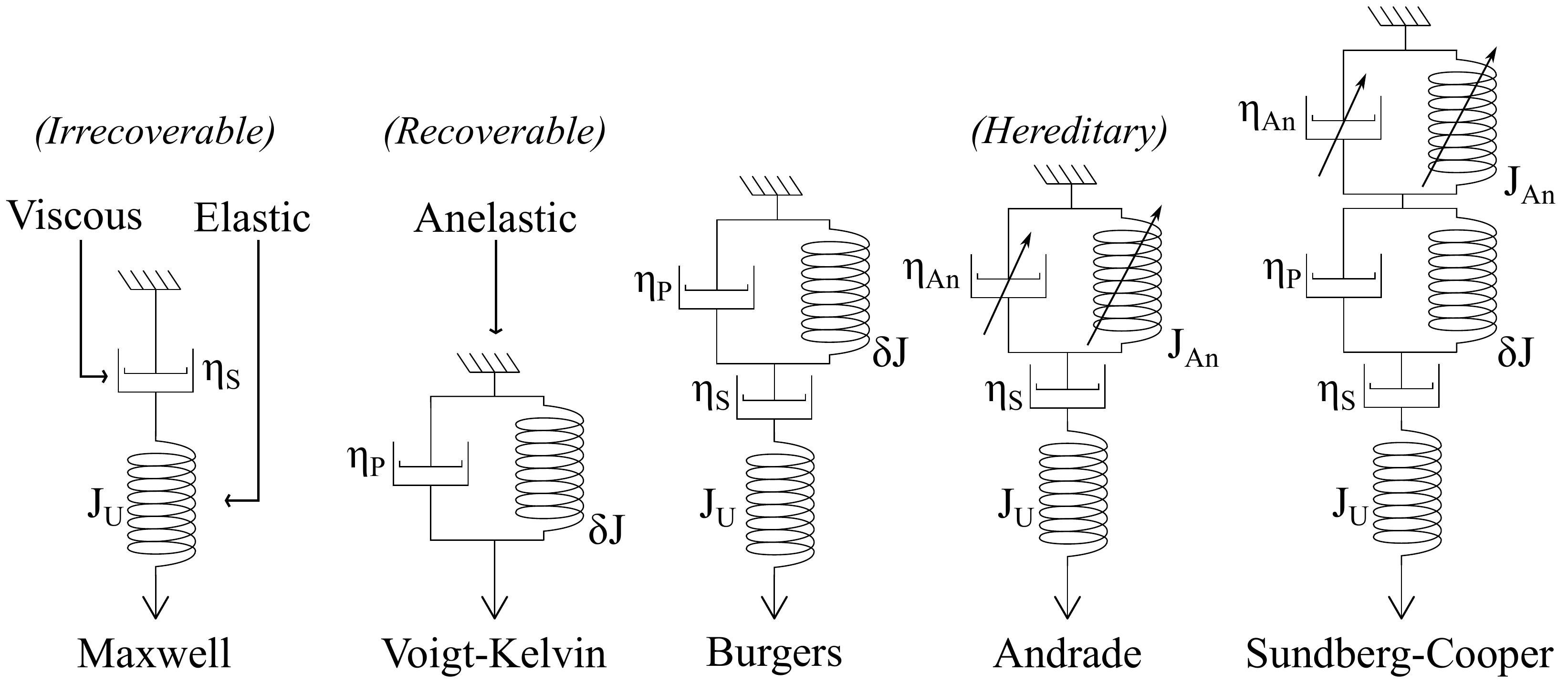}
    \label{fig:spring_dashpot}
    \caption{Representations  of the rheological models used in this study. A spring (with compliance $J$) represents an element that exhibits purely elastic rigidity. A dashpot (with viscosity $\eta$) is an element that exhibits purely viscous damping. $\maxcomp$ and $\voigtcomp$ respectively represent the `unrelaxed' and `defect' compliances (see Table \ref{tbl:parameters} for values). The unrelaxed compliance measures the strength of a material immediately after a stress load is applied. The defect compliance is defined as the difference between the relaxed and unrelaxed compliances, $\voigtcomp = J_{R} - \maxcomp$, where the relaxed compliance is a measurement at infinite time after application of load. The two viscosity terms $\maxvisc$ and $\voigtvisc$ are determined by the dominant creep viscosity. Elements marked by $\eta_{An}$ and $J_{An}$} depict the hereditary Andrade mechanism, which is contained within both the Andrade rheology and Sundberg--Cooper rheology. The varistor-like symbology reflects these elements modeling a broadened response spectrum.
\end{figure*}

Applied tidal theory has in the past been dominated by the use of two models. First, particularly within the field of extrasolar planets (following methods originally matured for analysis of binary stars), it is customary to use what we refer to as the fixed quality factor model, or \textit{fixed-$Q$} model. This model has no rheological underpinning, and simply uses a scalar-valued $Q$ factor, combined with the body's static Love numbers, to characterize all dissipative processes within a planetary object. As most often used, a fixed-$Q$ approach neglects any frequency dependence of the response (or does so by testing a small range of values), and relies upon selecting $Q$ values that have been confirmed through observation among solar system objects with similar characteristics (typically radius, mass, or density) to the object under study. This method, however, is highly susceptible to major errors, due first to the strong frequency dependence of most microscale dissipation mechanisms, and second to the fact that major differences in internal temperature and partial melt composition may often exist for planets of similar outward bulk properties \citep{Henning2009, HenningHurford2014}. It has also been observed that forcing frequencies change on astronomical timescales \citep{MD, HussmannSpohn2004}; so, while it remains very useful for first-round analysis, the use of a fixed-$Q$ for time domain studies will fall short in describing a planet with changing orbital and interior conditions. \par

The next step in complexity is the use of the Maxwell rheology, which has seen widespread use for tidal studies within our solar system \citep[e.g.,][]{RossSchubert1986}. The Maxwell model considers an element of rock or ice to consist of a perfect mechanical spring in series with a perfect mechanical damper (or ``dashpot,'' see Figure \ref{fig:spring_dashpot}). In concert, these elements create a material that, upon loading, experiences instantaneous elastic deformation, followed by unlimited viscous relaxation. A sinusoidal applied load leads to a damped and phase-lagged sinusoidal response. The Maxwell model captures some of the role of frequency dependence in planetary dissipation, but in general turns out to have a dependence that is too strong in comparison to real materials, and lacking in important subtleties such as regions in the frequency domain where a response temporally flattens. \par

Using the Maxwell model as a baseline, we compare three other rheological models (see Figure \ref{fig:spring_dashpot}) that have the potential to generate large tidal responses in regimes that are traditionally thought to be tidally quiescent. All of these models are characterized by an instantaneous elastic response, followed by some form of viscoelastic damping. Each pairing of spring and damper in a mathematical model leads to a characteristic frequency (analogous to \textit{RC} circuits in electrical engineering), at which the material will generally experience a peak response, both in amplitude and in energy loss rate. These may be thought of as forms of material resonance, akin to a classical harmonic oscillator. For the Maxwell model the corresponding period for its material resonance frequency, or Maxwell time, can be calculated as $\eta{}J$ using the material's viscosity, $\eta$, and compliance, $J$ (inverse of shear rigidity, $J=M^{-1}$). \par

All rheological models are attempts to represent the microphysical interactions between atoms and grains of a planet's bulk material on a macroscale, typically with a compact set of equations. Most models have been developed to match basic viscous and/or elastic responses, or to match specific datasets. Later attempts to associate such models with specific grain-scale phenomena have had mixed success \citep[see discussion in][]{McCarthyCastillo-Rogez2013}. However, we present some overarching comments on the specific rheological models used in this study, all of which have some degree of consensus in the material science community. \par


The Burgers rheology \citep{Peltier1986, Yuen1986, Sabadini1987,FaulJackson2005} is able to better capture certain interface interactions at grain boundaries. These become relevant at moderately high frequencies and are generally described by a peak or plateau in response. Grain boundary slip is a phenomenon that occurs on a shorter relaxation timescale than Maxwell-like diffusion creep, and is furthermore recoverable, as represented by the parallel spring--dashpot (Voigt--Kelvin) element pair within Burgers. This recoverable \textit{anelastic} strain is unique to rheological models that possess a transition between a fully elastic response and a viscous one. The Burgers model also contains a Maxwell element that represents classical diffusion creep, where non-recoverable motion is thought to occur through vacancy migration inside of grains. Such diffusional creep dominates at high temperatures and/or low frequencies. Studies of Postglacial rebound in particular have suggested that the Burgers body may be a more appropriate model of Earth's upper mantle than a Maxwell body, although perhaps over a limited range of temperatures and frequencies. Using parameters suggested by Earth-based observations \citep[see][]{Henning2009} leads to a rheological response in the temperature domain that is similar to Maxwell except at temperatures in the range 1200--1600 K, where a modest secondary peak in tidal dissipation occurs. The Burgers model is often extended by the inclusion of multiple peaks (each described by a different parallel spring--dashpot pair as seen in Figure \ref{fig:spring_dashpot}, added in series). The particular peaks included are generally chosen to fit specific datasets, and are not able describe higher frequency attenuations. \par


The Andrade model was originally developed to describe the strain response in laboratory samples of copper metal \citep{Andrade1910}. It has since expanded to become particularly successful in describing a broad range of laboratory studies, including silicate minerals, metals, and ices, and has recently made its way into planetary science. \par

One feature of the Andrade rheology is the goal of `softening' the too-steep frequency dependence of the Maxwell model with a function that is a power law in the frequency domain, with fractional powers of $\omega$ less than 1. The Andrade model is similar to another valuable concept in material science, that of a response plateau, also sometimes referred to as an attenuation band. Such a plateau is visible in the frequency domain for the applied-stress version of a behavior, and represents a material achieving a very similar level of attenuation over a broad range of frequencies. This is in sharp contrast with the Maxwell model, where peak attenuation occurs at one mathematically exact frequency, with a sharp fall-off on either side. Such a peak takes the form of a Debye peak \citep{NowickBerry1972}, which is visually similar to the more familiar Gaussian curve. Shifting models away from mathematically exact attenuation peaks has been referred to as ``response broadening,'' and the Andrade model exhibits features of such a useful shift. This is achieved in the model by considering not a spring and dashpot with conventional pure single-valued parameters, but instead a model where the elements include integration over a continuum of spring constants and damping coefficients. This in effect allows the model to incorporate the very real phenomenon that few real-world materials are composed of exactly one grain size; they typically contain impurities along with a spatially varying range of defects and defect densities. Response broadening has been attributed, at least in part, to such grain scale diversity, but the exact reasons for it do remain in discussion. Perhaps most importantly is the Andrade model's embrace of \textit{hereditary reaction}. Such a reaction is different from a purely viscous response whose details are lost after load is removed (irreversible). A hereditary reaction retains some aspect of material `memory' (which can be either reversible and irreversible) \citep{Efroimsky2012b}. This memory is dependent not just upon static material properties (as the Voigt--Kelvin model is), but also on how the aforementioned microphysical properties have changed with time.\par


Presented in \citet{SundbergCooper2010} as a better fit to laboratory data is a series combination of an Andrade mechanism with a Burgers rheology. \citet{SundbergCooper2010} discovered in their experiments on high-temperature olivine that a Burgers-like attenuation peak tended to appear in conjunction with a background attenuation best characterized by the Andrade model. As neither the Burgers nor Andrade formalism was able to fit this feature, they developed a composite rheological model blending features of both. We refer to their composite model here as the \textit{Sundberg--Cooper rheology}. The experiments of \citet{SundbergCooper2010} are of particular value to the planetary community, in that they were conducted both with useful mantle-analog material samples and at mantle relevant temperatures. The samples used were peridotite, primarily composed of olivine with the remainder (39\% by volume) composed of orthopyroxine, with characteristic grain sizes of around 5 $\mu$m. Temperatures tested ranged from $1473$ to $1573$ K. Although the experiments were conducted at 1 atm pressure, high-pressure work remains rare, and temperature has consistently proven to be the most critical environmental parameter in determining a material's bulk viscoelastic behavior, at least within one phase. In seeking the most relevant rheological extensions beyond Andrade to test, we find the Sundberg--Cooper model the most useful, in contrast to the somewhat ad hoc extended Burgers models, whereby response broadening is achieved more arduously via the piecemeal addition of single-resonance-frequency spring--damper pairs. Furthermore, the composite model presented by \citet{SundbergCooper2010} has features that make it likely to be as useful and fundamental as predecessors such as Maxwell, Andrade, and Burgers. For instance, the secondary attenuation peak in the Burgers subcomponent can be modified to fit various microphysical processes, while keeping the attenuation flatting provided by the Andrade subcomponent. \par

Even more material response models exist for materials relevant to a terrestrial planet's interior, including the rheologies of \citet{Lomnitz1956}, \citet{Becker1925}, and \citet{Michelson1917}. Even more are discussed in the context of ices by \citet[][and references therein]{McCarthyCastillo-Rogez2013}. A large proportion of these other models arise from empirical functions developed to fit early laboratory data. Many of these models have not seen widespread adoption for simple reasons, such as the fact that differing mathematical formulations lead to results that are not especially unique, such as the close comparison between the Lomnitz rheology and the Becker rheology \citep{MainardiSpada2012, StrickMainardi1982}. In other cases, models such as the Michelson rheology \citep[e.g.,][]{Lomnitz1956} contain a very large number of empirical coefficients, which are designed to improve a fit to one set of laboratory data, but which do not link back especially well to specific microcrystalline properties or phenomena. A general rheology model, such as the one presented by \citet{Birger1998}, shows promise in switching between these different models based on strains, temperatures, and forcing frequencies. However, the Andrade and Sundberg--Cooper rheologies are deemed here to be modestly superior test cases in that they first encompass the basic laboratory results that the Lomnitz and Becker rheologies were also created to capture (that of response broadening across a much wider range of input frequencies than a Maxwell model, also known as quasi-frequency independence), yet have the additional advantage of being anchored by far more modern geophysical and laboratory experiments. \par

\citet{Birger2006} raises a number of issues for Earth's mantle rheology that advanced planetary modeling may eventually need to consider. At very high strain levels, the Andrade rheology may require further adjustments for when power-law creeping flow begins to occur. \citet{Birger2012} states that a rough numerical threshold for this transition may occur at a strain of $10^{-3}$--10$^{-2}$. Strain within Io depends on the assumed rigidity, location, and time within an orbit, but falls typically in the range 1--3$\times$10$^{-6}$, as determined in tests using the methods of \citet{HenningHurford2014} or more simply by Equation 4.192 of \citet{MD}. For very short-period Earth-mass exoplanets some strain terms may reach 1$\times$10$^{-5}$--5$\times$10$^{-4}$, raising the possibility of local flow regions entering into this transition zone, given that Birger notes that mantle convection stresses can locally alter the dominant creep mechanism. Rheological anisotropies can also exist even in a single mantle-relevant crystal, even ahead of considering a polycrystal matrix. Given that even lateral temperature inhomogeneities in a convecting mantle cannot yet be considered in most present tidal methods \citep[excepting, perhaps, techniques such as][]{Sotin2002, Frouard2016}, these points serve as a reminder of the magnitude of work required to eventually unite modern material science with the modeling of other worlds.
\subsection{Compressibility and Tidal Magnitude Uncertainty}\label{sec:compressibility}

The model discussed in Section \ref{sec:background} assumes that the bulk of a planet is incompressible. This assumption will begin to break down for objects that have large interior pressures due to higher masses. The threshold where incompressibility is no longer valid is dependent upon composition, differentiation, and heat flux \citep[see Section 10.7 in][]{Schubert2001}. Our understanding of compressibility within the Earth is not yet complete. It has been suggested that compression effects will be localized rather than global in an Earth-sized body \citep{Schubert2001, Behounkova2010}. Whether or not this extends to larger exoplanets is still up for debate, but recent work suggests that compressibility will matter \citep{LiuZhong2013, Cizkova2017a}. Other work has indicated that compressibility may be important in certain materials within much smaller worlds, such as high-pressure ices within Ganymede \citep[][and references therein.]{NeveuRhoden2017}. \par

Compressibility may alter the thermal evolution of a large planet in two primary ways. First, compressibility (and pressure in general) will alter some thermodynamic parameters that are major inputs to our model. The pressure dependence of these parameters has had increased attention in both laboratory studies and theoretical modeling. Density tends to have the strongest dependence, and for the Earth this effect leads to an approximately 65\% increase in density at the core--mantle boundary (CMB) \citep{Schubert2001}. Thermal expansivity and specific heat both decrease with increasing pressure, although the most dramatic changes happen when $P<150$ GPa \citep[see Figure 1 in][]{Cizkova2017a}. In general, \citet{Cizkova2017a} found the pressure dependence of these parameters to suppress the vigor of convection and increase the effective viscosity of the mantle. \citet{LiuZhong2013} found similar results that were dependent upon the heat fluxes across thermal boundary layers. The full implications of these works on the long-term thermal state of a planet will require further study. We speculate that a reduction in convective vigor due to compression may introduce some fascinating scenarios where a mantle would be better able to retain heat while also being a weaker dissipater of tidal energy due to the increased effective viscosity. Such scenarios should be considered in future work when pressure and temperature dependence of thermodynamic parameters are better understood. In this work we are more concerned with the dependence of rheology on thermodynamic parameters. We implicitly model pressure-induced changes to some parameters by looking at phase spaces such as that for viscosity (Figure \ref{fig:shear_visc}). \par

Perhaps most significant to the questions we address here is the influence of compressibility on tidal dissipation itself. Equation \ref{eq:tidal_work} below is derived from the assumption that a planet is incompressible. Indeed, tidal studies that assume compressibility are greatly lacking in the literature, with the work of \citet{Tobie2005a} being a notable exception. A full derivation of the response of a compressible planet may be found in Appendix A of \citet{SabadiniVermeersen2004}, and this is compared to the incompressible (multilayer) response matrix of  Equation A3 in \citet{HenningHurford2014}. The considerable number of Earth-sized and larger exoplanets that appear to be in tidally active systems warrants a robust exploration of compressible tidal models. This is an area that we plan to explore in future work when we incorporate multi-layer solutions \citep{SabadiniVermeersen2004, HenningHurford2014, NeveuDesch2015}. \par 

For this article we continue to use an assumption of incompressiblility to explore large extrasolar planets. One defense of this approach is grounded in our interest in the morphology of dissipation, rather than specific magnitudes. We do not anticipate the overall shape of dissipation (over the domains of interest) to greatly change when transitioning into a compressible regime. Likewise, since compressibility will modify all rheological models, the comparison between models presented throughout Section \ref{sec:results:exoplanets} is still valid. Finally, prior work finds that tidal dissipation is often strongest at shallow depths where alterations in outcome due to compressibility are weakest \citep{HenningHurford2014}. For silicate worlds near or greater than the mass of the Earth, tidal heating presumably concentrates very strongly into any shallow, low-viscosity asthenosphere (in a frequency-dependent manner), and the relative tidal response of all lower layers is often small. If such low-viscosity upper layers are common, this could help mitigate the concern of using an adjustment for compressibility for worlds of super-Earth mass, because the primary driver of the tidal outcome in such cases would become the thickness and viscosity of any asthenosphere. The same argument applies for worlds with an ice shell atop a silicate core, where tidal activity concentrates strongly into the ice at all typical planetary forcing frequencies. \par

Due to the paucity of compressible models used for tides both for the solid Earth and in Earth-analog exoplanets, the degree of error that any compressible correction may induce is not clear. However, it is well established for tidal heating that uncertainty in the selection of viscosity-determining parameters (setpoint viscosities, activation energies) overwhelmingly dominates uncertainty in tidal heat production. Note that the pressure dependence of viscosity on Earth, as modeled in Arrhenius laws by an activation volume term $V^*$, is itself subject to broad concern.  Determinations of the viscosity structure of Earth's mantle, to the depth of the CMB \citep[see][]{MitrovicaForte2004}, find viscosities bounded in the range $10^{20}$--$10^{24}$ Pa s, with non-monotonic trends. Use of almost any surface-relevant estimate of activation volume $V^*$ (see value range in Section 7.6 of \citet{TurcotteSchubert2002}) in a pressure-dependent silicate viscosity law leads to divergences from this structure by many orders of magnitude (e.g., CMB viscosities near $10^{30}$--$10^{36}$ Pa s). See Figure 1 and Section 3.3 of \citet{HenningHurford2014} for a more complete discussion. Therefore, a robust predictive model of high-pressure silicate viscosity is still lacking, even for the Earth, and this governs tidal outcomes more than anything else. This exemplifies the point that attempts to predict the exact magnitude of tidal exoplanet outputs are in their infancy, and parametric uncertainties that lead to changes of say $\sim$ 5\%--10\% in dissipation are still dwarfed by uncertainties of multiple orders of magnitude from other sources. As demonstrated below, the choice between the Andrade and Maxwell models is exactly one such larger-scale correction that can lead to 10--100$\times$ corrections. It is not yet known if the alpha and zeta parameters of the Andrade and Sundberg--Cooper rheologies vary significantly with pressure or density. \par

\section{Methods}\label{sec:methods}
To perform comparisons between rheological models, we first focus our study on a single generic planetary system. Then, in Section \ref{sec:results:exoplanets}, we explore implications for certain extrasolar systems. To provide context to results we look at an Io-like satellite orbiting a Jupiter-mass host (see Table \ref{tbl:parameters} for planetary and orbital parameters). We assume that this satellite is subjected to forced eccentricities, much like Io is held in an eccentric orbit due to the Laplace resonance between Jupiter and the other Galilean moons. However, to simplify the interpretation of discrete thermal phenomena in time, we merely apply external eccentricity patterns such as step functions and sine waves, instead of explicitly modeling the orbits of any other satellites. \par

\subsection{Interior and Thermal Models}\label{sec:thermal_model}
Following methods similar to recent studies of tidally active bodies \citep[e.g.,][]{HussmannSpohn2004, Henning2009, ShojiKurita2014}, we track the average temperature of the satellite's mantle, $T_{m}$,

\begin{equation} \label{eq:mantle_temp}
\dot{T}_{m} = \frac{\dot{E}_{Radio}+\dot{E}_{Tidal}+Q_{CMB}-Q_{Conv}}{\left(\text{St} + 1\right)M_{m}c_{m}},
\end{equation}

and core, $T_{c}$,

\begin{equation} \label{eq:core_temp}
\dot{T}_{c} = \frac{-Q_{CMB}}{M_{c}c_{c}},
\end{equation}

over time. The Stefan number, \textit{St}, is defined by using the latent heat of the mantle ($L_{m}= 3.2\times10^{5}$ J K$^{-1}$) as \citep{ShojiKurita2014},

\begin{equation} \label{eq:stefan}
\text{St} = \frac{L_{m}}{c_{m}(T_{l}-T_{s})}.
\end{equation}

 This average mantle temperature is used to calculate the mantle's effective viscosity and compliance (the inverse of rigidity). $Q_{CMB}$ is the heat passing through the core--mantle boundary. $Q_{Conv}$ is the total heat escaping the mantle due to convection. $M_c$, $M_m$, $c_c$, and $c_m$ are the masses and specific heats of the core and mantle respectively. The mantle is heated by the decay of radiogenic isotopes, $\dot{E}_{Radio}$. For both Io and exoplanets, we assume radiogenic rates for silicate material that match the modern bulk silicate rate on Earth, assuming Earth's current Urey ratio is 0.5 \citep{Jaupart2007TemperaturesEarth}. This allows even scaling of radiogenic outputs by mass. Unless otherwise stated, radiogenic rates are varied backwards in time, after partitioning into major isotope contributions and accounting for each individual half-life. \par 
 
 Tidal heating within the homogeneous and incompressible mantle, $\dot{E}_{Tidal}$, is given by \citet{Segatz1988},

\begin{equation}\label{eq:tidal_work}
\dot{E}_{Tidal} = -\text{Im}\left(k_{2}\right)\frac{21}{2}\frac{\left(R_{sec}n\right)^{5}}{G}e^{2}f_{tvf},
\end{equation}

 and related to the forced eccentricity $e$, orbital mean motion $n$, and the rheological response described by $-Im(k_{2})$, the imaginary portion of the second-order Love number \citep{Love1892, PealeCassen1978, Segatz1988, Efroimsky2012a}\footnote{Equation \ref{eq:tidal_work} is valid for low eccentricities, zero inclination, and synchronous orbits. For more information see \citet{MakarovEfroimsky2014}}. Tidal heating is expected to be focused within the mantle and not the core \citep{HenningHurford2014}. Equation \ref{eq:tidal_work} accounts for this with the scaling factor $f_{tvf} = V_{mantle}/V_{planet}$ for the tidal volume fraction \citep{Henning2009}. This represents the volume fraction in active tidal participation, given that three of the five powers of $R_{sec}$ in Equation \ref{eq:tidal_work} arise from a linear dependence on an object's total spherical volume during the derivation of the homogeneous tidal equation \citep[see][]{MD}. This serves as a rough approximation of the true multilayered behavior of a tidal system \citep[e.g.,][]{TakeuchiSaito1962, SabadiniVermeersen2004, Tobie2005, RobertsNimmo2008, Wahr2009, JaraOrueVermeersen2011, HenningHurford2014}. The negligible tidal output of the core is the most significant difference between a homogeneous tidal model and a multilayer model, followed by the presence or absence of an asthenosphere. Lithospheres for silicate systems are also in general too cold to contribute significantly to tidal activity, which is additionally captured in the use of $V_{mantle}$ above, even though lithosphere volumes are small. Note that replacing $V_{mantle}$ with $V_{asthenosphere}$ would effectively convert Equation \ref{eq:tidal_work} into a useful approximation for a multilayered world that contains an asthenosphere, given that asthenosphereic tidal heating strongly dominates when present. Such approximate corrections are linear in Equation \ref{eq:tidal_work}. This is most effective when dominant layers are thick, such that layer bending is not an issue, as arises for the ice shell of Europa. \par

Heat is assumed to be transported out of the core into the mantle, and later out of the mantle to the surface by convection separated by conducting boundary layers. We use a parameterized macroscale convection model that utilizes thermal boundary layers at the top and bottom of the mantle \citep[][and references therein]{OConnellHager1980, ShojiKurita2014}. The thickness of the mantle's upper boundary layer $\delta_{upper}$ is found as 

\begin{equation}\label{eq:blt_upper}
 \delta_{upper} = \frac{D_{m}}{2a}\left(\text{Ra}_{c}\frac{\eta\kappa}{\left(T_{m}-T_{surf}\right)\alpha_{V}gD_{m}^{3}}\right)^{1/3},
 \end{equation}
 
in terms of the mantle's critical Rayleigh number $\text{Ra}_{c}$, mantle thickness $D_m$, surface temperature $T_{surf}$, and further terms defined in Table \ref{tbl:parameters}. The lower boundary layer of the mantle $\delta_{lower}$ can be related to the upper boundary layer if one assumes a fixed increase in viscosity from top to bottom \citep{NimmoStevenson2000, ShojiKurita2014},
 
 \begin{equation}\label{eq:blt_lower}
 \delta_{lower} = \frac{\delta_{upper}}{2}\left(\gamma\left(T_{c}-T_{m}\right)\right)^{-1/3}exp\left(\frac{-\gamma\left(T_{c}-T_{m}\right)}{6}\right),
\end{equation}
 
with $\gamma$ representing the increase in viscosity. The heat escaping both the core and mantle is limited by conduction through these boundary layers,
 
\begin{equation}\label{eq:q_conv}
Q_{Conv} = 4\pi{}R_{m}^{2}k_{m}\frac{T_{m}-T_{surf}}{\delta_{upper}},
\end{equation}

\begin{equation}\label{eq:q_cmb}
Q_{CMB} = 4\pi{}R_{c}^{2}k_{m}\frac{T_{c}-T_{m}}{\delta_{lower}},
\end{equation}

where $k_m$ is the mantle thermal conductivity, and $R_c$ and $R_m$ the outer radii of both the core and mantle. Note that a thermal boundary layer is an inescapable result of a convective system due to the turning trajectory of convective material. Because not all material in the flow pattern is able to make direct contact with the layer above (or below), the heat from any given parcel of material is forced to move via conduction through the last small distance of the convective layer. The thickness of this boundary layer has been empirically related to the vigor of convection via the Rayleigh number. Material in a thermal boundary layer is moving with the convective flow, and is not the same as a stagnant lid wherein all horizontal movement has ceased. We assume no stagnant lid. A full time evolution model will require the creation of a stagnant lid when internal heat flux is sufficiently low as to create a thick, strong conductive barrier to near-surface horizontal deformation. If thermal equilibrium is assumed, it is theoretically possible, but would remain to be seen by future modeling, that a stagnant lid with very efficient heat-pipe penetration could offer low thermal resistance, but perhaps only in rare circumstances. Mantle convection would still proceed below such a lid for long durations, and heat-pipe activity passing through even a thick lid would still be allowed. Detailed entry into and exit from such states is a complication that should be addressed in future models.\par

The surface temperature of the satellite may be approximated by assuming that graybody radiation from the surface is sufficiently rapid to match diurnally averaged insolation heating and the total heat coming from the interior, as characterized by the instantaneous convective cooling rate, 

\begin{equation}\label{eq:surf_temp}
T_{surf} = \left(\frac{\left(1-A\right)L_{*}}{16\pi{}a^{2}_{*}\epsilon_{v}\sigma_{B}} + \frac{Q_{Conv}}{\sigma_{B}}\right)^{1/4}.
\end{equation}

Here $L_{*}$ is the stellar luminosity, $a_{*}$ the stellar distance, $\epsilon_v$ the emissivity, and $\sigma_B$ the Stefan--Boltzmann constant. This assumption of radiant equilibrium is not the same as overall thermal equilibrium, and allows heat production within the world to vary away from the current convective cooling rate. We also assume a thin/minimal atmosphere with no significant greenhouse effect.  \par

\citet{FischerSpohn1990}, later expanded by \citet{Moore2003}, described a range of tidal-convective equilibrium states, whereby the total radiogenic and tidal heat production rate for Io (or any similar world) is matched by the bulk rate of convective cooling. Convective cooling rises monotonically with temperature, with the slope increasing sharply at the onset of melting, due to falling bulk viscosity and rigidity. Note that this model, like all parameterized convection models, is based on averaged behavior, and sudden bursts or lulls of convective activity, as well as local variations, are possible for real systems. As can be seen in Figure \ref{fig:heating_temp}, tidal heating as a function of temperature typically includes one or more peak values, leading to a range of opportunities for the total heating and cooling curves to cross. Both stable and unstable equilibrium states are possible at these crossing locations, where energy in equals energy out. The stability of a given crossing may be determined by considering perturbations from the exact value. If, for example, heating exceeds cooling on the low-temperature side of equilibrium, then the temperature is naturally restored from the perturbation, leading to stability. \par

Tidal-convective equilibrium systems typically contain a hot stable equilibrium (HSE) just after $T_{br}$, the breakdown temperature\footnote{For minerals, the breakdown temperature, or disaggregation temperature, is the point in partial melting where solid grains lose mutual contact in a growing fluid bath, above which a material rapidly takes on bulk properties more resembling a fluid.} (which we assume to be around 1800 K for peridotite at Io pressures \citep{Moore2003}). A cold-unstable equilibrium typically exists well below the solidus temperature $T_s$. Systems evolving in time will be attracted toward stable equilibrium points, and repelled from unstable points, with relatively little time spent in between. Because it induces a second low-temperature peak in tidal dissipation, the Burgers rheology has the unique opportunity to express two pairs of both stable and unstable equilibrium points \citep{Henning2009}. Tidal-convective stable equilibrium points are typically extremely stable due to the steep slope of both tidal and convective cooling curves in the onset-melting region where they often meet. Note that meeting in this region is in large part a function of forcing frequency, and thus the typicality of this description reflects the typical nature of studying both moons and exoplanets with orbital periods in the range 1--20 day. The location of equilibrium points is also a strong function of orbital eccentricity. See \citet{Henning2009} for bifurcation diagrams describing how stable and unstable equilibria evolve with varying $e$. Similar diagrams could readily be constructed where semimajor axis is the term controlling total tidal magnitude (such as when inward or outward migration is induced by external non-tidal phenomena). For any given system, we also expect a critical eccentricity, below which tidal heating is so weak that no tidal-convective equilibrium points exist. Such equilibrium states are essential for understanding the time evolution of tidal-convective systems, which we explore in Section \ref{sec:results:time}. \par

Heat-pipe activity \citep[e.g.,][]{Moore2001} causes the vigor of cooling to rise even more sharply when a system is heated just a few per cent, by melt fraction, beyond the solidus. While the convection-only cooling curve rises to a near vertical slope at the breakdown temperature, a system with advection has its cooling curve rise to near vertical approximately 1--3\% above the solidus. This generally acts to shift the HSE point from near $T_{br}$, to near $T_s$ (assuming homogeneous behavior). This location is often below typical maximum viscoelastic tidal heating rates. But the relative slope of the heating and cooling functions remains such that, even in the case of heat-pipe activity, the HSE point is strongly stable. We do not linger on this issue, because the HSE value is very similar across all rheologies described here, and this convective/advective difference has been described previously for a Maxwell response. \par

\subsection{Dependence of Material Strength on Temperature and Partial Melting}\label{sec:partial_melt}
We allow the mantle's homogeneous material to melt based on fixed solidus and liquidus temperatures (respectively, 1600 and 2000 K). These values are calculated for olivine at Io's mid-mantle pressure of $\sim1.5$ GPa \citep{Takahashi1990}. The strength and effective viscosity of the mantle will depend upon both the temperature and melt fraction. We assume that the viscosity will decrease with increasing temperature via an Arrhenius relationship. The rate of decrease will become rapid once a critical melt fraction (50\%, corresponding to the breakdown temperature) is reached, eventually becoming that of a liquid once the mantle is completely molten. Likewise, the strength of the mantle will decrease at this critical fraction \citep{MooreHussmann2009}. The strength and effective viscosity affect both the convective vigor of the mantle and the rheological response. See Sections 4.2 and 4.3 in \citet{Henning2009} for all equations required to define this melting behavior of viscosity and shear modulus in detail. We use the medium-strength case of the three models given there.\par

\subsection{Rheological Response}\label{sec:rheo_response}
The imaginary part of the second-order Love number, used to calculate the tidal heating within the mantle, is found via the compliance of the mantle \citep{Efroimsky2012a},

\begin{equation}\label{eq:imk2}
-\text{Im}\left(k_{2}\right) = -\frac{3}{2}\frac{J_{U}\tilde{\mu}\text{Im}\left(\bar{J}\right)}{\left[\text{Im}\left(\bar{J}\right)\right]^{2}+\left[\text{Re}\left(\bar{J}\right) + J_{U}\tilde{\mu}\right]^{2}},
\end{equation}

where $\bar{J}$ is the complex compliance, or creep function, of the mantle. The functional form of $\bar{J}$ for each rheology we consider is given in Table \ref{tbl:creep_functions}. $J_{U}$ is the unrelaxed compliance, and $\tilde{\mu}$ is the effective rigidity---a measure of the relative strength of a planet relative to its own gravity. Equation \ref{eq:imk2} is derived from the definition of the static Love number, $k_{2} = (3/2)(1 + \tilde{\mu})^{-1}$ \citep{Love1892}, once recast in the complex form, $\bar{k}_{2} = (3/2)(1 + \tilde{\mu}J_{U}/\bar{J})^{-1}$. We follow the notation of the classic text of \citet{NowickBerry1972} where $M$'s denote rigidities (specifically for tides, shear moduli), and $J$'s denote their inverse. Here $\bar{J} = \bar{M}^{-1}$ just as the static compliance $J = M^{-1}$. The algebraic similarities between the static and complex Love numbers, compliances, and rigidities are due to the \textit{correspondence principle} \citep[see Section 4 in][]{Efroimsky2012b}\footnote{As mentioned in \citet{Efroimsky2012b}, the formalism presented here is general only to the extent that the correspondence principle holds. Adjustments will be needed for tides caused by librations in longitude due to any triaxiality of the tidal body \citep{FrouardEfroimsky2017}. We also do not consider apsoidal or nodal precessions \citep{EfroimskyMakarov2014}.}. For reference, we have derived the equations for $-\text{Im}\left(k_{2}\right)$ for both the Andrade and Sundberg--Cooper models (Table \ref{tbl:love_numbers}), and written them in terms of the fundamental element parameters that are visualized in Figure \ref{fig:spring_dashpot}. It may be more convenient to use the real and imaginary components of the complex rigidity in a particular simulation suite, so we also provide those derivations in Tables \ref{tbl:rigidity_function:table_a} and \ref{tbl:rigidity_function:table_b}.\par

The phase angle, $\epsilon_{2}$, by which strain differs from applied stress can be expressed in a similar form \citep{Efroimsky2012a},

\begin{equation}\label{eq:lag}
\tan\left(\epsilon_{2}\right) = -\frac{\maxcomp\tilde{\mu}\text{Im}\left(\bar{J}\right)}{\left[\text{Im}\left(\bar{J}\right)\right]^{2} + \left[\text{Re}\left(\bar{J}\right)\right]^{2} + \maxcomp\tilde{\mu}\text{Re}\left(\bar{J}\right)}.
\end{equation}

\citet{BiersonNimmo2016} performed a thorough analysis comparing Io's measured $\text{Im}(k_{2})$ to a predicted value using a reduced Andrade model. It is important to understand when their assumptions, made to reduce the general Andrade formula, are applicable. They correctly point out three different regimes for the Andrade $\text{Im}(k_{2})$ value \citep[see Eqns. 17--19 in][]{BiersonNimmo2016}, and state that Io is likely to fall into the following constraints (adapted from their notation to ours\footnote{Most earlier work on Andrade uses the parameters $\beta$ (proportionality parameter) and $\alpha$ (exponent parameter). Instead of $\beta$, we follow the reasoning of \citet{Efroimsky2012a} and use the $\zeta$ first defined in that work. $\beta$ has mixed dimensions that in turn depend upon $\alpha$. This creates additional conceptual confusion when presented with various values of $\beta$. In contrast, $\zeta$ has a physical meaning (albeit an enigmatic one): the ratio of the characteristic Andrade timescale to the traditional Maxwell one. Other nomenclature exists for $\alpha$ as well, but it is generally interchangeable, with the exception pointed out by \citet[][Section 3.4]{Efroimsky2012a}. We address the frequency dependence stated in that work in our Section \ref{sec:andrade_freq_dependence}}), first: $ \left(\andcomp\andvisc\omega\zeta\right)^{-\alpha}\alpha! \ll \tilde{\mu}$, and second: $\left(\andcomp\andvisc\omega\zeta\right)^{-\alpha}\alpha! \gg 1$. In the case of Io with the nominal compliance and viscosity values found in Table \ref{tbl:parameters}, along with $\alpha = 1/3$, these assumptions approximate to (a) $5\times10^{-3}\zeta^{-1/3} \ll 50 $, and (b) $5\times10^{-3}\zeta^{-1/3} \gg 1$. We note the following warnings for those who wish to apply this version of the Andrade model to situations beyond the scope of \citet{BiersonNimmo2016}. These two conditions create an opposing constraint on $\zeta$ with little room for error. For example, if we choose the nominal value of $\zeta = 1$, then condition (a) is satisfied while condition (b) is not. \citet{BiersonNimmo2016} note the experimental work of \citet{Jackson2002} and \citet{Jackson2004} who found $\beta\sim10^{-13}$--$10^{-11}$ Pa$^{-1}$ s$^{-1/3}$ which corresponds to $\zeta\sim10^{-10}$--$10^{-4}$. Choosing a middle value of $\zeta=10^{-8}$ we find that both conditions are achieved, but only just. Since both viscosity and shear modulus are included in these formulae, any changes in temperature and/or melt will dramatically affect the results (for example, as major morphological alterations to Figure \ref{fig:shear_visc} below). \par

Beyond these concerns, it should also be noted that a reduced model will need to be modified whenever a system crosses the aforementioned regimes. It may be easy to miss a crossing, especially in the case of exoplanets with effective rigidities that are lower than Io's, which will further constrain the above assumptions. For instance, the ratio $\tilde{\mu}/\mu$ is about five times larger for Io than for the median TRAPPIST-1 planet where $\tilde{\mu} = 1.52\times10^{-10}\mu$ compared to Io's $8.23\times10^{-10}\mu$ \citep{Gillon2017SevenTRAPPIST-1, Wang2017UpdatedPlanets}. Lastly, this logic locks a material parameter ($\zeta$) to system-specific characteristics. In all likelihood, $\zeta$ will vary as a function of pressure, temperature, and forcing frequency within a non-homogenized planet. In the end, we recommend the use of the general Andrade model (see Table \ref{tbl:creep_functions}) for all but the most constrained questions. \par

\subsection{Andrade Parameters and their Frequency Dependence} \label{sec:andrade_freq_dependence}
The Andrade exponent, $\alpha$, has been constrained between 0.1 and 0.4 \citep{WeertmanWeertman1975, GribbCooper1998, Jackson2002} for olivine with slightly lower values for other rocky/icy materials \citep[][]{McCarthy2007, McCarthyCastillo-Rogez2013}. We explore a range of different $\alpha$ values to account for this uncertainty. $\zeta$, is defined as the ratio between the Andrade and Maxwell characteristic timescales, $\zeta = \tau_{A}/\tau_{M}$ \citep{Efroimsky2012a}. The value of $\zeta$ is determined by the underlying creep mechanisms compared to a purely Maxwellian creep. We assume that diffusional creep is dominating within Io's mantle \citep{AshbyVerrall1977}. Under diffusional creep $\tau_{A}\sim\tau_{M}$, thus we expect $\zeta\sim1$ \citep{WebbJackson2003, Castillo-Rogez2011}. This assumption can fall apart in many interesting tidal cases, such as for exoplanets where pressures may change the dominant creep mechanism. Some laboratory studies on Earth materials have found $\zeta$ to be quite small ($10^{-10} < \zeta < 10^{-4}$ \citep{Jackson2002, Jackson2008}). \citet{Jackson2004} also found\footnote{$\zeta$ calculated from their $\beta\sim1\times10^{-2}$ using the viscosity and compliance values of the partial melt.} values of $\zeta\sim1$. \par

The Andrade anelasticity, in both the pure Andrade model and as a subcomponent of the Sundberg--Cooper model, is suspected to reduce to a Maxwell-like viscoelasticity below a critical frequency \citep[see discussions in][]{Efroimsky2012a, Efroimsky2015}. This is expected since any transient effects governed by the Andrade hereditary terms will be dominated by slow, viscous dissipation at low frequencies. Below this critical frequency it is believed that the jamming/unjamming of dislocations, grain boundary sliding, or some combination of both will cause this anelastic-to-viscoelastic transition \citep{KaratoSpetzler1990,Miguel2002}. It has been suggested \citep[e.g.,][]{Birger2006} that a Lomnitz rheology is better suited at these low frequencies, but at different strain levels. In the end, a general model may require many rheological components to account for these dependences. The complexities of analyzing such models are difficult given the uncertainties in each rheological model's parameters. Instead of wading through these nuances, we examine a mantle that is subjected to a single rheology no matter what its temperature or frequency. However, to account for a potential low-frequency cut-off, we compare a static Andrade rheology to one in which the Andrade timescale parameter, $\zeta$, is allowed to increase exponentially below a cut-off of $\omega_{crit}\sim1$ day$^{-1}$ as

\begin{equation}\label{eq:zeta_freq}
\zeta\left(\omega\right) = \zeta_{0}\exp\left(\frac{\omega_{crit}}{\omega}\right).
\end{equation}

A large $\zeta$ will cause the Andrade response to reduce to that of Maxwell, as can be seen in its creep function. The critical frequency is in turn dependent upon temperature and the activation energy(ies) of the underlying mechanisms \citep{KaratoSpetzler1990}. Its value could be much larger than the one considered in this work (for example $\omega_{crit}\sim1$ yr$^{-1}$ in \citet{KaratoSpetzler1990}). Rather than modeling the temperature dependence of $\omega_{crit}$, we set its value to be something applicable for the system under study (Io's orbital period is 1.7 days) for comparison purposes. We implicitly explore other possibilities by manually changing $\zeta$ (as well as $\alpha$) independently of $\omega_{crit}$ in Section \ref{sec:results:andrade}. \par

\section{Results} \label{sec:results}
\subsection{Equilibrium Results} \label{sec:equilibrium}
\begin{figure}[hbt!]
    \centering
    \label{fig:heating_temp}
    \includegraphics[width=1\linewidth]{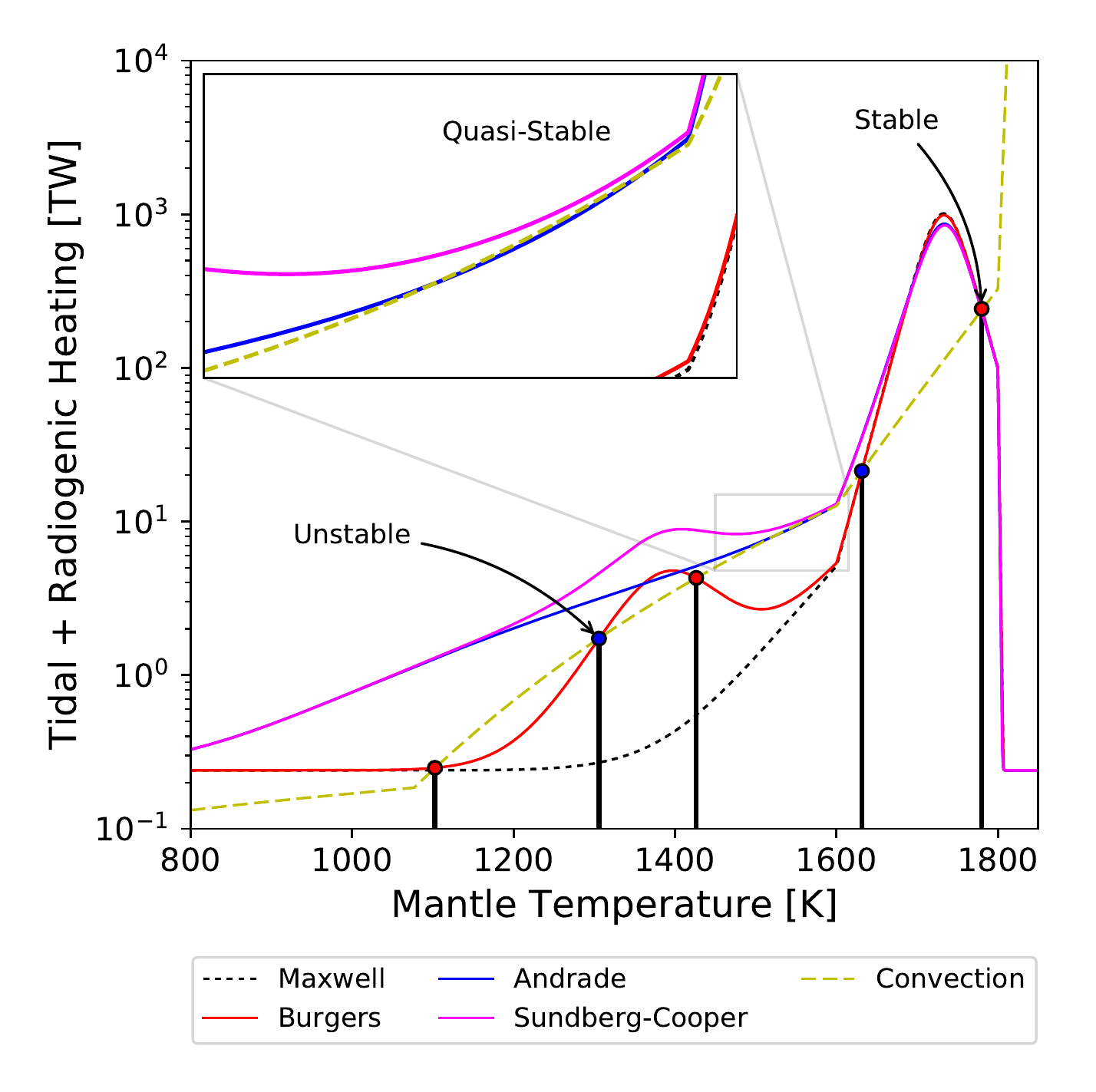}
    \caption{Total rates of mantle heat production summing tides and radionuclides are plotted against mantle temperature for the four rheologies studied, in an Io-analog setting with an assumed solidus of 1600 K, $T_{br}$ = 1800 K, and mantle shear modulus $\maxshear = \maxcomp^{-1} = 60\times$10$^{9}$ Pa. To illustrate a full set of possible equilibria, all heating curves use half Io's modern $e$. Convective cooling as a function of temperature is shown in dashed yellow. Crossover points between convective cooling and total heating indicate equilibrium points (both stable and unstable) discussed in Section \ref{sec:results:time}. The shallow slope of the Andrade and Sundberg--Cooper models allows (at half  Io's modern eccentricity) the emergence of previously unreported tidal-convective equilibrium category: a quasi-stable region of temperature $\sim$ 500 K wide. Deviations between Maxwell and the other rheologies occur mainly in the range 1100--1600 K. The position and magnitude of the secondary Burgers peak seen in both the Burgers and Sundberg--Cooper models, occurring at $T\sim1400$ K, is sensitive to our choice of $\voigtshear\coloneqq\voigtcomp^{-1} = 5\maxshear$ and $\voigtvisc = 0.02\maxvisc$}.
\end{figure}

Equilibrium states form when convective cooling is approximately equal to internal heat generation, shown as dots in Figure \ref{fig:heating_temp}. Depending upon the thermal--orbital conditions and rheology, a planet could have multiple equilibrium points. These points will also vary over time as a satellite's orbit changes \citep[e.g.,][]{OjakangasStevenson1989,FischerSpohn1990,Saxena2018}. Both convection and tidal heating are functions of temperature and partial melting. Crossover points that fall on the right side of a peak in heating (red filled circles in Figure \ref{fig:heating_temp}) are considered to be stable equilibria. If the mantle temperature increases or decreases from these points, then the heating or cooling acts to drive the temperature back into equilibrium. Crossover points on the left slope of a heating peak are unstable (blue filled circles) and mark the divide between recoverable (to the right of unstable points) and unrecoverable mantle temperatures. Here a `recoverable' mantle is defined as one that is able to maintain high tidal dissipation at a given fixed eccentricity, with a mantle at, or trending toward, a stable equilibrium. In Figure \ref{fig:heating_temp}, all rheological models have effectively the same HSE before the mantle breakdown temperature ($T_{br}\approx1800$ K). If a mantle reaches this equilibrium then it will be able to maintain high temperatures (with large melt fractions) for long time periods, assuming the forcing eccentricity is not significantly dissipated. The Burgers rheology produces a secondary peak to the left of the primary Maxwell peak due to its secondary material resonance. This leads to the possibility of additional equilibrium positions. This secondary peak allows a mantle to maintain a moderate temperature (with near zero melt fraction) for long time periods. A similar secondary peak occurs for the Sundberg--Cooper model; however, for the value of $e$ in Figure \ref{fig:heating_temp} there is no crossing with convection as occurs for the weaker Burgers curve. \par

The position and amplitude of any secondary material response peak due to the Burgers mechanism are determined by the choice of parameter values for the Burgers (parallel spring--dashpot) element, either in the Burgers model itself or imbedded within the Sundberg--Cooper model. The peak location is determined akin to the position of the Maxwell peak, but via a relaxation timescale arising from $\tau_{Burg} = \voigtvisc\voigtcomp$, just as Maxwell time is defined as $\tau_{Max} = \maxvisc\maxcomp$. In the temperature domain, the peak then occurs when $\voigtvisc (T, \omega)$ causes $\tau_{Burg}$ to match the forcing period. The choices of $\voigtvisc$, $\voigtcomp$ (and its equally relevant activation energy) are poorly constrained \citep[see Section 4.4 for discussion][]{Henning2009}. However, modest perturbations from the selected values leave the system behaviors described here intact, because the Burger's peak continues to allow secondary equilibrium points across a wide range of positions/amplitudes. The main change in outcome would occur if future measurements find that preferred values for the Burgers element are so close to terms for the Maxwell element that the Burgers and Maxwell peaks combine into one, in which case the complex behaviors inherent in possible low-temperature equilibria would vanish. Currently, such blending is not considered likely based on existing laboratory data. The amplitude of the Burgers peak is also influenced by astrometric terms such as planet size, as discussed in Section \ref{sec:results:str_visc}. \par

Interestingly, the Andrade subcomponent produces a shallow-sloped decay of dissipation with dropping temperature. In the inset plot of Figure \ref{fig:heating_temp} we see that the Arrhenius-controlled convection produces an overlap for a range of temperatures in the rheologies with an Andrade subcomponent. In the example shown, tidal heating is larger than convection on both sides of this region. The end result will be a slow increase in temperature throughout this quasi-equilibrium before a quick jump to the HSE (this can be seen in Row 3 of Figure \ref{fig:laplace:eccen}). While there may be a mathematical point where the actual crossover between heating and cooling occurs, the importance of any such exact point is debatable in a real object experiencing latitudinal, longitudinal, and temporal deviations from averaged behavior. This region, however, introduces a new type of equilibrium that Andrade-controlled mantles could exhibit at moderate temperatures. Emergence of this $\sim$ 500 K wide feature requires only a mild reduction from Io's modern forcing, at half Io's present value of $e$, alongside center-of-range Andrade mineralogical terms. This subtle overlap will depend upon the relative strength of convection vs. tidal heating. A shifting eccentricity (as investigated in Section \ref{sec:results:time}) can cause Io, or any exomoon analog, to spontaneously slip into or out of this quasi-equilibrium band. Io's magma eruption temperatures \citep[see][]{Keszthelyi2007a, Davies2011} are compatible with large portions of Io's mantle being in this broad quasi-stable equilibrium position today. This could suggest a lower $e$ in Io's recent past, or merely be coincidental. More likely is the possibility of a tidal-advective HSE point near $T_s$ = 1600 K at the modern $e$ = 0.0041.\par

\subsection{Strength and Viscosity} \label{sec:results:str_visc}
\begin{figure*}[hbt!]
    \centering
    \label{fig:shear_visc}
    \includegraphics[width=.8\linewidth]{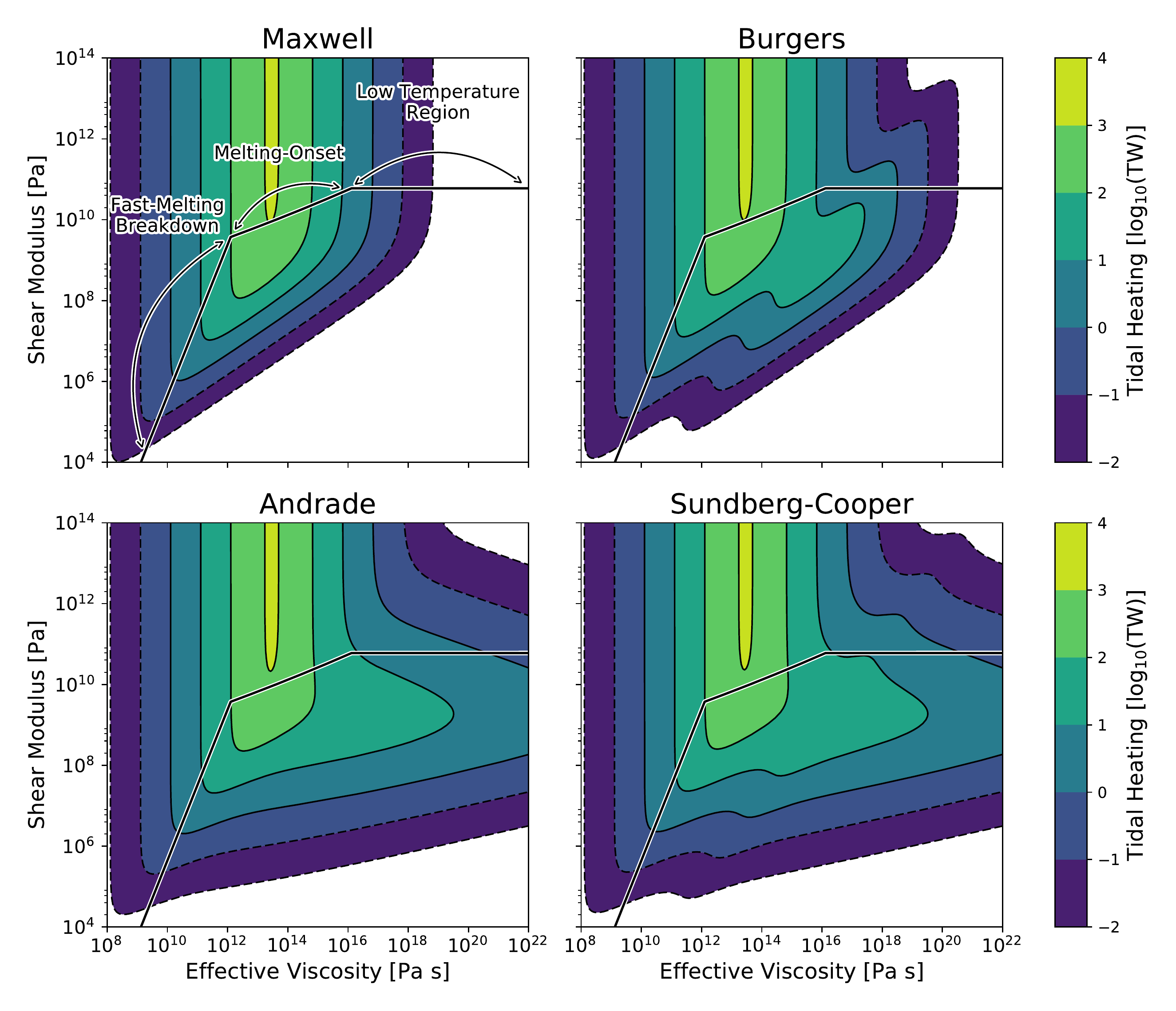}
    \caption{Tidal heating rate is mapped as a contour in the phase space of shear modulus vs. effective viscosity. Right and left sides of individual plots respectively represent cool and warm mantle temperatures, relative to the mantle's melting point. The solid line represents a typical path that a planetary object could take as it melts (leftward along the trajectory) or crystallizes (rightward). The tidal heating is given for Io with its present-day semimajor axis and $e = 0.5e_{present}$. Rheological models with increasing complexity, starting from the Maxwell model (upper left) to the Sundberg--Cooper model (lower right) express a trend toward increasing the range of both parameters over which elevated tidal dissipation will occur. Note how evolutionary trajectories pass through the regions of enhanced tidal activity of the Andrade and Sundberg--Cooper cases in the high-viscosity regime. This is the primary point that makes these rheologies highly relevant for this system.}
\end{figure*}

To assess the behavior of the Andrade and Sundberg--Cooper rheologies relative to other rheological models we look at phase space maps of shear modulus plotted against a mantle's effective viscosity (Figure \ref{fig:shear_visc}). Such a phase space is useful for visualizing how and why the tidal dissipation of a planetary object varies during the process of melting or crystallization. The map for the Maxwell rheology is well documented \citep{Segatz1988, FischerSpohn1990}, and contains a single `ridge' of high tidal dissipation, which attenuates as one approaches low values of shear modulus. A typical trajectory for a planetary mantle undergoing melting in such a map (white and black line in Figure \ref{fig:shear_visc}) is to begin on the far right side (cold, high viscosity). As a mantle warms, viscosity decreases rapidly, but the shear modulus remains constant so long as the temperature is well below the solidus. Once near or above the solidus temperature, modest shear weakening begins. For forcing frequencies akin to Io's of around 1--10 days, a melting trajectory typically crosses the Maxwell-like ridge during this weakening phase. \citet{Henning2009} describe the existence of a separate `island' of dissipation that occurs for the Burgers rheology. Depending on the Burgers parameters, the forcing frequency, and most importantly the mass \citep{Henning2010} of the planet, the position of this secondary island may shift such that the melting trajectory may either directly cross it or miss it entirely. This determines the extent to which Burgers-like behavior is relevant for a given orbital scenario. \par

The Andrade subcomponent (found both in pure Andrade and in Sundberg--Cooper) produces a spectrum of shear modulus and viscosity values that together lead to greater overall energy dissipation \citep{ShojiKurita2014}. This spectrum is restricted to cooler temperatures, but is very broad and encompasses many different combinations of mantle states. In the shear-viscosity phase space of Figure \ref{fig:shear_visc}, this appears as a blurring of the Maxwell-like high-dissipation ridge, extending to much higher viscosities. This blurred region is partly akin to the Burgers island, in that it occurs in a similar region and accomplishes a similar outcome: increasing the parametric region within which moderate tidal dissipation may occur. Similar to the isolated Burgers island, expression of this Andrade region for a given world's time evolution is sensitive to the value of the initial (or final) cold-state shear modulus. If the value is high, less of the Andrade-like broadening will be experienced. This implies that Andrade will be especially important for cold brittle ice mantles, with lower shear moduli ($\sim4\times10^{9}$ Pa) than silicate shear moduli ($\sim$5--6$\times10^{10}$ Pa). \par

Like the Burgers model, the Sundberg--Cooper rheology also contains a localized and elevated response ``island''; however, in this case the island is more significantly joined to the Maxwell ridge by the overall response broadening of the simultaneous Andrade-like activity. In this way, the shear-viscosity map for Sundberg--Cooper is satisfyingly what may be expected to arise from a linear combination of its precursor elements, expressing all the features of each. It is also therefore subject to the same principles as Burgers and Andrade alone, in terms of the ability for particular trajectories to either hit or miss its unique features, as well as the manner by which a planet or moon's total mass helps to control the vertical position of the high-dissipation features relative to a given fixed parametric trajectory. Unlike Burgers, however, Sundberg--Cooper reduces such sensitivity significantly, and thus ameliorates the concern that the selection of exact Burgers terms constitutes something of a mathematical idealization. \par

\begin{figure*}[hbt!]
    \centering
    \label{fig:mass_tuning}
    \includegraphics[width=0.8\linewidth]{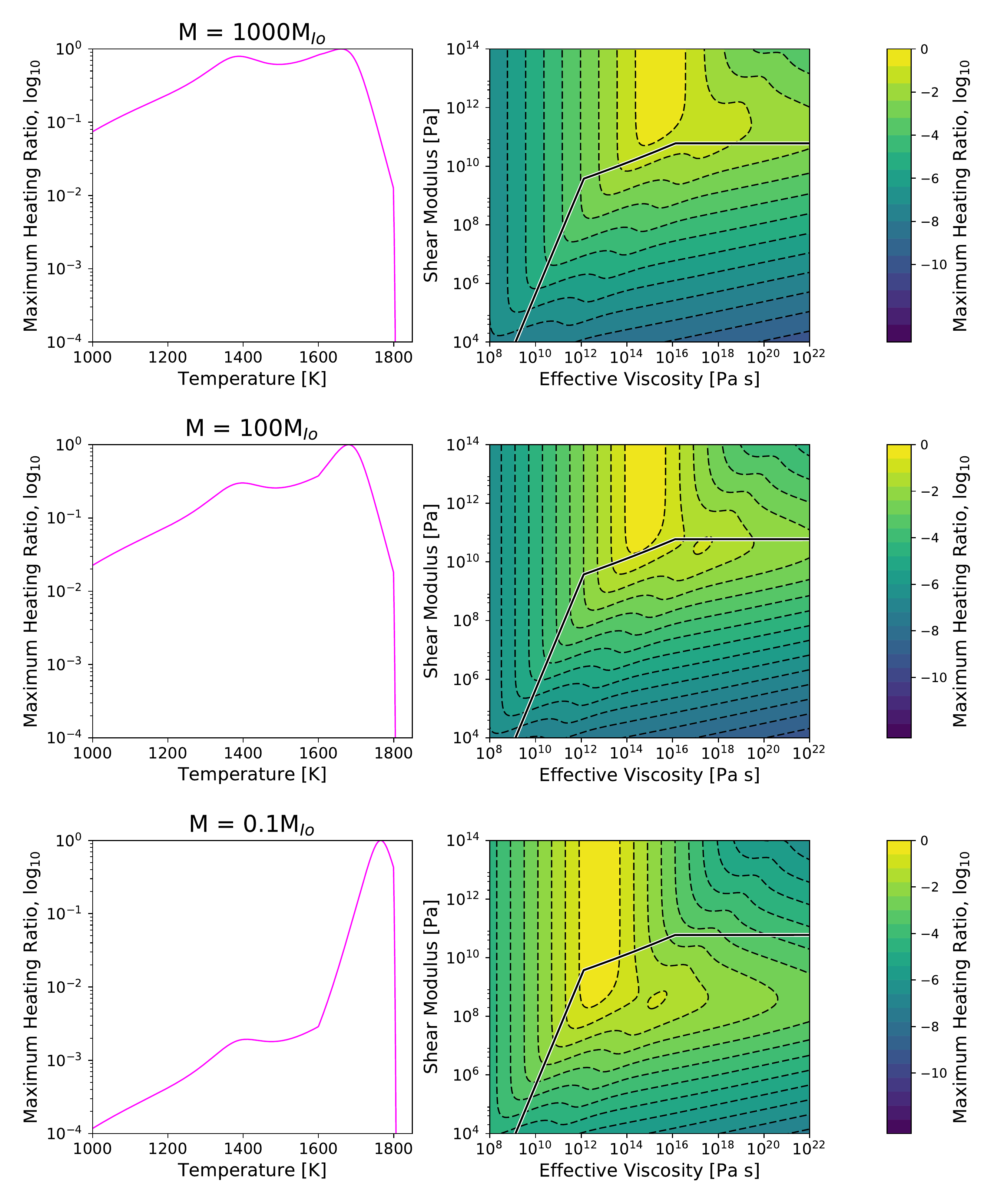}
    \caption{Planetary mass is the primary control on which regions of tidal phase space an object experiences. We find that the mass of the secondary in which tides are being generated is the main control on the vertical positioning of the underlying structure from Figure \ref{fig:shear_visc}. Nominal unmelted shear modulus may shift the horizontal position of the melting trajectory, but only by small amounts, because plausible mantle shear moduli do not vary by as many orders of magnitude as object mass may. Objects much larger than, or much smaller than Io will not experience as many Andrade, Burgers, or Sundberg--Cooper dissipation features, and are thus better approximated by a Maxwell model. The unique structure of the Sundberg--Cooper rheology is most expressed at 100 M$_{Io}$ ($\sim$ 1.5 M$_E$).}
\end{figure*}

Figure \ref{fig:mass_tuning} demonstrates how the mass of the object in which tides are being generated, $M_{sec}$, uniquely controls the extent to which Burgers, Andrade, and Sundberg--Cooper features are expressed. Other parameters such as forcing frequency, semimajor axis, and perturber mass have no such role. Secondary mass exerts this control through the Love number. Alterations in $M_{sec}$, relative to a fixed (unmelted) shear modulus, in effect vary the extent to which the object dominated by gravity or by strength. Subsolidus changes in shear modulus have the same effect but cannot plausibly vary by the same order of magnitude. For any given choice of mineralogical parameters, there is thus an optimal $M_{sec}$ at which non-Maxwell features most prominently emerge. Such emergence takes two forms: the size of any other peaks besides the high-temperature Maxwell peak, and the amount of elevation of the low-temperature tidal background. For our model, such optimal tuning occurs at $\sim100$ M$_{Io}$ (about 50\% more massive than Earth). The notable relevance of non-Maxwell features continues up to 1000 M$_{Io}$, and down to 0.1 M$_{Io}$. 

One of the most important basic principles in Figures \ref{fig:shear_visc} and \ref{fig:mass_tuning}, climbing up from Maxwell to Sundberg--Cooper, is the steady expansion of high-dissipation regions, reflecting the inclusion of more and more diverse grain-scale phenomena as gained through the steadily improving empirical match of each model to laboratory results. \par

Recall from Section \ref{sec:compressibility} that we utilize tidal equations derived with an assumption of incompressibility, as well as with parameters such as $\alpha$ that are not modeled as varying with pressure. Larger solid exoplanets are exactly the venue within which it may be most important for tidal research to steadily evolve to including compressible cases, despite the cost of added mathematical complexity. The impact of compressibility on tidal heat magnitudes for worlds in the range 1--10 $M_{E}$ cannot be known until such studies are carried out. The impact may be either large or small, but the key is the necessity to be aware of the assumption, and use that awareness to guide future research. We highlight that the effects discussed in this section will be valid even for a compressible planet: the mass tuning is due to the gravity and radius dependence of the effective rigidity, a term that is still present in the compressible derivation of dissipation \citep{SabadiniVermeersen2004}.

\subsection{Time Domain} \label{sec:results:time}

Figure \ref{fig:heating_temp} informs us that the Burgers, Andrade, and Sundberg--Cooper rheologies will have the greatest impact for cooler mantles. This implies that as an object secularly cools from a hot state, it may pass through many points where tidal dissipation is enhanced compared to a Maxwell model. In the time domain, we test a range of behaviors to explore changes this may cause both for generic systems as well as uniquely for Io. \par

\begin{figure*}[hbt!]
    \centering
    \label{fig:step_func}
    \includegraphics[width=1\linewidth]{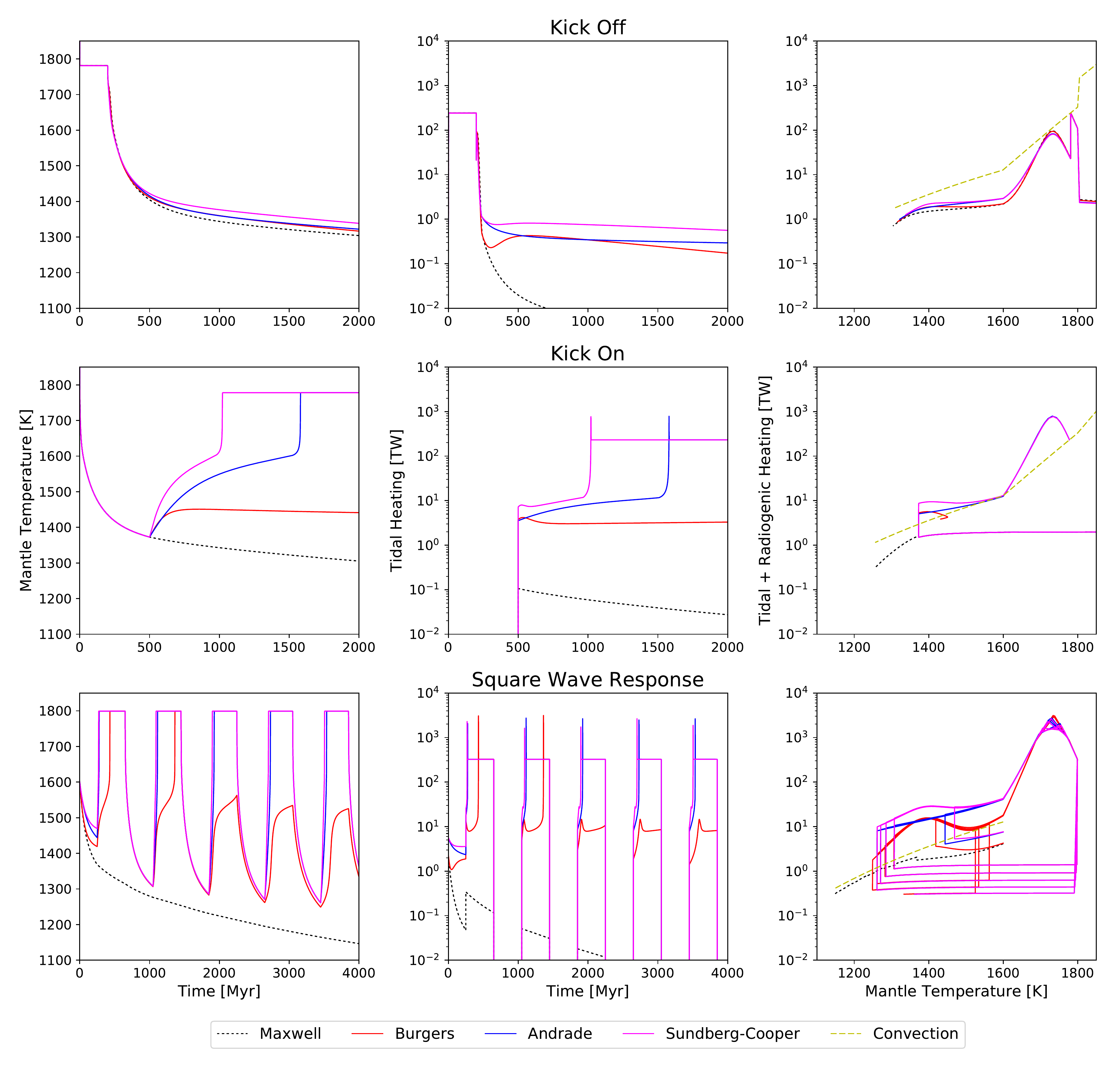}
    \caption{We demonstrate the individual response of each rheological model to a sudden loss of eccentricity (Row 1), a gain of eccentricity (Row 2), and a continuous loss/gain (modeled by a square wave, Row 3). When a non-zero eccentricity is imparted to the secondary, its dissipation will move into equilibrium with convective cooling (Column 3). Depending upon the temperature at ``kick-on,'' a rheology may or may not find its HSE. Even if a rheology finds its HSE it may only be on the border of losing HSE due to any perturbation. The continuous loss of radiogenic heating may push a mantle over this border (see Burgers rheology in Row 3 and Figure \ref{fig:exoplanet_time}).}
\end{figure*}

First, consider a step response to a change in tidal forcing. Such a change may occur due to a variation in eccentricity or semimajor axis. A step response is physically possible in the form of an orbital scattering event such as a three-body encounter, but here we simply wish to use it to understand the basis of more complex orbital behaviors to come next.  In Figure \ref{fig:step_func}, we show (Row 1) how an Io-like moon would respond to both a sudden decrease in tidal forcing (using a drop in eccentricity from $e=0.55e_{present}$ to $e=0.16e_{present}$) and a sudden increase (Row 2, $e=0$ to $e=0.75e_{present}$). The step-down response shows that both Andrade and Sundberg--Cooper lose their temperatures slightly more slowly than a Maxwell body. Likewise, for an upward step, both models warm the object faster. In fact, if secular cooling has proceeded too long, some rheologies may not respond to the upward step at all, faced with mantles that have become too viscoelastically cold. Parameters in Figure \ref{fig:step_func} are chosen to show a case where Maxwell is unable to respond but other rheologies can. Depending on the parameters chosen, the secondary peak in the Burgers and Sundberg--Cooper models may either be transiently expressed in an upward step event or may even be settled upon as a new equilibrium (as in the Burgers case does in Figure \ref{fig:step_func}). \par

Changes in Io's eccentricity, mean motion, and consequently heating rate depend strongly on Jupiter's $Q$ value, which does not appear explicitly in our model, because we are testing the response of an Io-analog to simplified step functions and sine functions in eccentricity that are exactly applied. $Q$ of Jupiter mainly controls how much power is extracted from Jupiter's rotational energy by Io (through tides) and is thus transferred into the resonance-locked satellite system. This action is essential to the long-term stability of the Laplace resonance, because dissipation in Io tends to evolve the system away from exact resonance (inward migration away from Europa), while dissipation in Jupiter drives the system back toward exact resonance (migration of Io toward Europa). Whether the system is in equilibrium between these effects has been a longstanding debate, and limits to the plausible range of $Q_J$ have likewise been a central component of Laplace resonance theory \citep[see, e.g.,][]{GoldreichSoter1966, Sinclair1975, Yoder1979, Greenberg1987}. Our model does not resolve these debates, but does add the need to also consider the perspective and limits of geological behavior in the debate. Our model is in essence a direct response to the results of \citet{HussmannSpohn2004}, in terms of the diversity of amplitude, shape, and period of oscillations in eccentricity that are possible in their fully coupled system. \citet{HussmannSpohn2004} use a value of $Q_J = 1.2\times10^5$. While the exact evolutionary histories that their model produces may change with variations in $Q_J$, the appearance of a diversity of resonance-induced oscillations is expected to be fundamental, both due to both orbital effects \citep[see for example][Section 8.9]{MD} and cyclic internal/geophysical changes in both Io and Europa (as additionally occur in \citet{HussmannSpohn2004}).

A step-response timescale (Row 3, Figure \ref{fig:step_func}) that allows full equilibration of interior temperatures before further changes is akin to a low-frequency square-wave response. Faster cycling leads to non-repeating behaviors. At high frequency, mantle temperatures may not move far from starting values before restoration of tidal forcing. This is true regardless of the depth of the change in forcing. However, at sufficiently low frequency, and with a sufficiently deep low excursion in eccentricity, a key phenomenon emerges (see Figure \ref{fig:step_func}, Row 2, Column 1). If a mantle is allowed to cool for long enough, it reaches a point from which, if $e$ is restored to its prior state, the tidal heating outcome does \textit{not} restore to the prior state for some rheologies. Instead, the mantle rock is too cool to respond, and despite the same restored forcing intensity, the rock viscoelastically fails to generate heat, and the world continues to cool. This effect can be exacerbated by the decay of radiogenic heating, which we explore further in Section \ref{sec:results:radio_loss}. \par

For models with multiple heating peaks such as Burgers and Sundberg--Cooper, the system may have complex opportunities to move between or be trapped in a range of tidal-convective equilibrium states. If the orbit keeps shifting, the thermal state may never reach full equilibrium, instead shifting with stable and unstable tidal-convective equilibria (themselves functions of eccentricity) acting as attractors and repellors. \par

\begin{figure*}[hbt!]
    \centering
    \label{fig:sine_func}
    \includegraphics[width=1\linewidth]{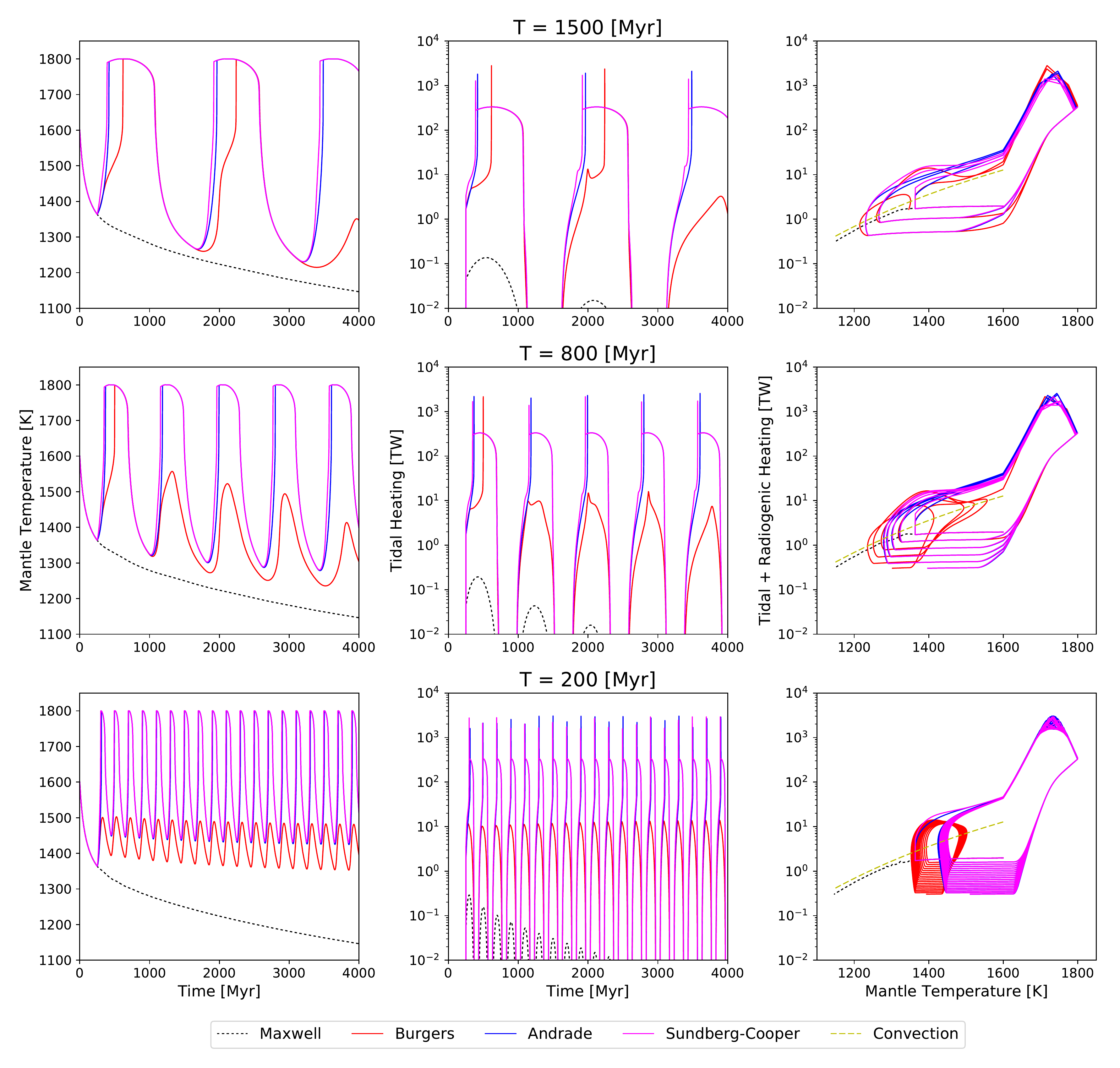}
    \caption{Three different sinusoidal eccentricities are imparted to an Io-like body. All Rows have the same eccentricity amplitude of $e=0.0025$. It is apparent that rheologies that take longer than a period to find their HSE will never find it. This can be seen in the first 1000 Myr for the Burgers rheology. It is able to find its HSE given enough time ($\sim300$ Myr) in Rows 1 and 2. However, it never finds it when the oscillation period falls below this (Row 3). The low oscillation period of $\sim200$ Myr matches those found in \citet{HussmannSpohn2004}. We again see a borderline-crossing effect in the Burgers rheology (Rows 1 and 2) due to radiogenic heating loss, first noted in Figure \ref{fig:step_func}.}
\end{figure*}

The rightmost column of Figure \ref{fig:step_func} shows the combined tidal and radiogenic heating of a system evolving in time. Curved trajectories, which look similar to Figure \ref{fig:heating_temp}, appear when the object is in a warming phase; however, when compared to Column 2, it can be seen that not all portions of the path are traversed at equal rates. Events such as material-resonance peak crossings can occur very rapidly. This plotting method becomes very useful for evaluating cyclic forcing, as in Row 3, Column 3, where the non-repeating nature of the response becomes evident. These also allow us to interpret how certain equilibrium points are (or are not) being crossed by an object. Such systems show a sensitivity to initial conditions akin to the hallmark deterministic non-periodic flow of classical dynamical models of chaos \citep{Lorenz1963}. We use this visualization in the rightmost columns of Figures \ref{fig:step_func}--\ref{fig:laplace:eccen}. \par

Figure \ref{fig:sine_func} next shows the response of this system to an applied sinusoidal variation in eccentricity. Rows 1--3 show the effect of varying the cycle period. Similarities in Column 3 to a Lorenz-style classical chaos attractor are even more pronounced in these cases. Sinusoidal variations in eccentricity are a standard outcome for systems locked in mean-motion resonances (MMRs) such as the Galilean moons. \citet{HussmannSpohn2004} showed typical oscillations in eccentricity for Io with periods of the order of 100--200 Myr, and amplitudes of $e \approx$ 0.001--0.003. Oscillations in semimajor axis are also standard for an MMR. Eccentricity and other orbital elements may also vary sinusoidally due to secular resonances \citep[][Sec. 8.5]{MD}. Both amplitude and period control internal thermal evolution outcomes, via control of a system's ability to approach and hold thermal equilibrium in concert with the orbital forcing. Andrade and Sundberg--Cooper systems generally have a far better ability to recover from low-eccentricity (or low forcing) excursions during a cycle, whereas Maxwell systems, if they become too cold, may pass below a threshold temperature for a given forcing intensity, from which they are unable to muster sufficient tidal activity to later recover on the upswing of a cycle. This may lead either to progressively slipping away from fully achieving the high-temperature tidal-convective equilibrium point at cycle peaks (see Maxwell and Burgers curves in Figure \ref{fig:sine_func}, Column 2) or simply failing to do so catastrophically in just one cycle (as did the Maxwell curve in Figure \ref{fig:step_func}, Row 2). Thus far more readily than its counterparts, a Maxwell simulation can become locked in a cold state from which it is unable to recover, despite tidal forcing being sufficient at the high point of the cycle to maintain tidal-convective equilibrium \textit{if} a mantle were already hot. This key difference in behaviors leads us to a range of conclusions for Io. \par

Let us introduce the term `tidal resilience' to mean a system's ability to maintain tidal activity in the face of perturbations, most notably via the orbital forcing. By this metric, Maxwell lacks tidal resilience compared to its alternatives. Low-$e$ perturbations can easily send Maxwell into an unchecked cooling pattern from which it cannot escape, unless $e$ is later pushed far higher than Io's modern value. The Andrade anelasticity within the Andrade and Sundberg--Cooper rheologies imparts both with excellent tidal resilience in contrast. Their low-temperature response is elevated, and this leads to far easier recovery from transient low-forcing states. \par

Observational evidence suggests that Io is at, or approaching, its hot stable tidal-convective (or tidal-advective)  equilibrium point \citep{Moore2003}. The very presence of melt and volcanism strongly suggests this, and the observation of some high-temperature magmas lends further support \citep{McEwen1998, Keszthelyi2007a, Davies2011}. The most credible upper limit is 1613 K \citep{Keszthelyi2007a}, which is a downward revision from estimates in \citet{McEwen1998}, due to nonlinear image movement across the CCD of the Voyager Infrared Interferometer Spectrometer and Radiometer. 50--100 K of alteration may occur from the interior, with an unknown balance of cooling due to adiabatic ascent, but also heating due to viscous dissipation in the magma column. Note that the HSE point for an advective (heat-pipe) dominated Io would occur only a few degrees above the solidus temperature, which we select as 1600 K, although compositional uncertainty and variation make this number substantially uncertain. But whether Io is at an HSE point or approaching it, the point is that the mantle is clearly not within the comparatively cold range of 1000--1300 K, the same range from which Maxwell has great difficulty escaping after any transient low-$e$ excursion. \par

If Io were best described by a Maxwell model, it would have far greater difficulty retaining this hot state for the $>$4 Gyr that Io has perhaps been in orbital resonance. Given that we believe Andrade or Sundberg--Cooper to be a better model of Io's mantle, we postulate that their resilience in the face of orbital forcing oscillations has been critical to the survival of Io's volcanoes. If a model such as Maxwell has ruled Io's silicate mantle, then one lengthy or large amplitude excursion of low eccentricity could have been sufficient to cool the moon far enough for tidal activity to never resume. Such a situation could have occurred prior to formation of the Laplace resonance, when eccentricity magnitudes were generally low overall. Alternatively, a perturbation may have occurred after the resonance was established and may have had the potential to break the resonance. The dramatic changes in eccentricity seen in the figures of \citet{HussmannSpohn2004} encourage us that such excursions are possible. Excursions in eccentricity may not even be necessary to invoke a low-temperature period within Io. A relatively quick cooling or melting phase within Europa and/or Ganymede's ice shell (part of the coupling architecture utilized by \citet{HussmannSpohn2004}) would dramatically change those bodies' dissipative response. This would impact the rate of change of Europa and Ganymede's mean motion, thereby influencing Io's orbital distance and tidal response. \par

As the inner Galilean moons are currently in the Laplace resonance, then either no resonance-breaking perturbation ever occured or Io was able to recover. Given the chaotic nature of the early Jovian system \citep[e.g.,][]{Hahn1999, Morbidelli2010} and the results present in \citet{HussmannSpohn2004}, we feel that the latter scenario is more likely. Therefore, Io's mantle may have cooled too much for the Maxwell model to recover (see the discussion related to their Figure 7). In that case, even if the orbits of the inner Galilean moons were able to return to their modern configurations, their interiors would have continued to cool. An alternative solution would require any such low dissipative event(s) to be paired with subsequent high dissipative event(s) intense enough to bring Io back out of a cold Maxwell-unresponsive state. We find that using realistic material models enables more low dissipative events and negates the need for high dissipative ones. The application of Andrade or Andrade-like rheologies may help to explain the mystery of how tidal activity on Io, once started, could have then continued uninterrupted for potentially billions of years despite a complex and ever-changing orbital environment. A counterargument to this could be given by some models that put Io closer to Jupiter in the past. A smaller separation distance would increase any rheology's ability to produce heat even with low forcing. Continued work on both the origin of the Laplace resonance and its evolution will be required to further address this question. \par

We note that fixed-$Q$ simulations in rocky bodies have the opposite shortcoming. They predict effortless continuity in tidal forcing, regardless of interior thermal evolution. They thus miss entirely the possibility of a body becoming too cold and failing to respond to tides. Fortunately, the most up-to-date material models achieve both orbital resilience and accuracy in one package. While our tests using prescribed step/sine functions of eccentricity may not include all complexity of a fully coupled tidal--orbital simulation, including freedom of the semimajor axis to vary, dissipation within the host, and behavioral associations to a host $Q$ value, they demonstrate how starting tidal activity from a cooler mantle is especially problematic for a Maxwell model.\par

\subsection{Implications for the Galilean Laplace Resonance} \label{sec:results:laplace}

An open question about the Jovian system is how long the Laplace resonance has been active \cite[][and references therein]{PealeLee2002}. Two top-level theories for the assembly of the Laplace resonance exist. In one, the moons migrate outwards \citep{Yoder1979, YoderPeale1981, Greenberg1987, Malhotra1991, Showman1997}, as they do now, under the influence of Jupiter's $J_{2}$ oblateness on $da$/$dt$. Early differences in the migration rate may plausibly allow moons that accrete in initially random locations to eventually cross their 2:1 MMR positions. Such crossings, if convergent, lead to locking into the resonance \citep{MD} and allow the moons to move in lock-step in order to link a third object into a 4:2:1 pattern. Alternatively, migration may occur inwards \citep{CanupWard2002, PealeLee2002, CanupWard2009}, as may analogously occur in exoplanet systems as Type I migration \citep[e.g.,][]{UdryMayorSantos2003, IdaLin2008}, due to magnetohydrodynamic torques induced by each moon within a primordial gas/dust disk out of which they have just formed. As is postulated for exoplanets, when the solar wind finally blows away the last of this accretion disk, inward migration ends and outward migration may begin based on Jupiter's $J_{2}$ value. While inward migration is occurring, it is possible for Ganymede to first sweep Europa into a 2:1 MMR, and then for the Europa--Ganymede assemblage to later sweep Io into the 4:2:1 final pattern seen today. \par

A key difference between these models is the timing. For inward migration, the Laplace resonance must form prior to loss of the debris/gas disk that induces inward movement. Unless such a debris disk formed late in Jupiter's history due to breakup of a prior moon or moon set, which is considered highly unlikely, this implies rapid assembly of the resonance pattern following Jovian accretion. It also implies that the Laplace resonance has been remarkably stable over time, precluding any dynamical perturbations sufficient to break it over the following $>$4 Gyr. Constraining the timing of the onset of the Laplace resonance by any alternative means may help to favor one model or another. The mechanism shown above, by which only certain rheologies allow for recovery from excursions with low eccentricity or low tidal forcing, provides us with one such new tool.  \par

\begin{figure*}[hbt!]
    \centering
    \label{fig:laplace:tau}
    \includegraphics[width=1\linewidth]{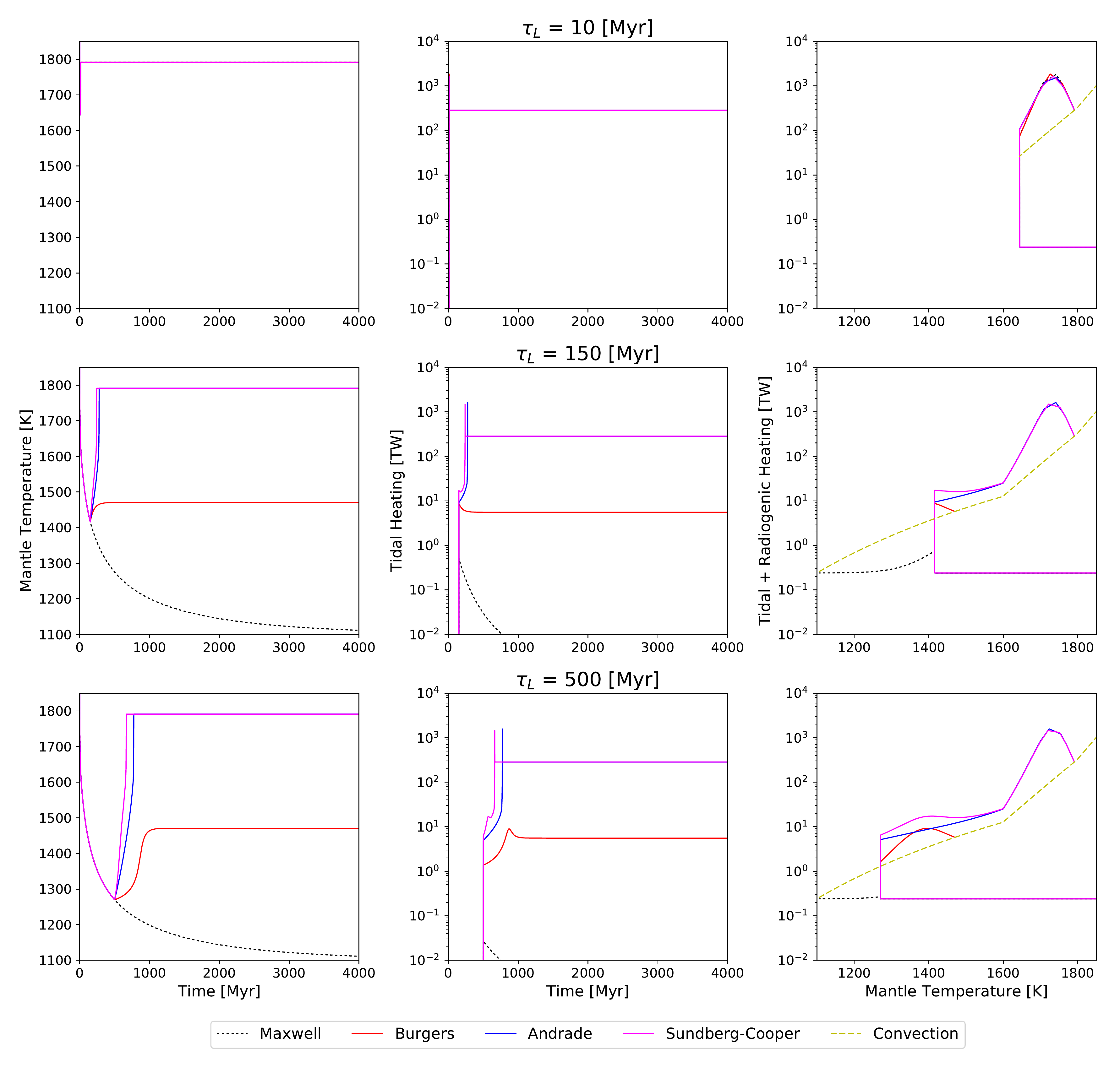}
    \caption{Io is assumed to have coalesced into a hot, differentiated, molten sphere at or just before $t=0$. After $\tau_{L}$ Myr, a long-duration, constant, forced eccentricity of $e=0.003$ is imparted to Io, mimicking the Laplace resonance that currently exists between the Galilean moons and Jupiter. Tidal dissipation, for multiple rheological models, will then counteract this eccentric orbit. Three different $\tau_{L}$ values are shown on three different rows. Column 1: average mantle temperature is shown as a function of time. Column 2: tidal heating is shown as a function of time. Column 3: tidal + radiogenic heating is plotted against the current mantle temperature. The last column is a useful way to visualize the position each rheology is at on the idealized Figure \ref{fig:heating_temp}. It also shows which equilibria are being reached, if any.}
\end{figure*}

Consider Io's first entrance into a tidally active state following its formation. If Io was formed in a circular orbit (e.g., prior to resonant forcing), or if any initial eccentricity quickly dissipated, then it would act as a secularly cooling sphere heated only by radiogenic decay (apart from gravitational energy released during early differentiation). When the Laplace resonance initialized it would impart a (likely varying) forced eccentricity on Io \citep[see Figure 5 in][]{HussmannSpohn2004}. If Io experienced significant cooling before this initialization, then a Maxwell rheology may not be able to return Io to a hot state due to its poor dissipation abilities at low temperatures. In Figure \ref{fig:laplace:tau} we test what effects realistic rheologies have on answering this question. For these results, we assume that Io coalesced at or just before $t=0$ and has a high internal temperature and melt fraction. We impose a forced eccentricity of $e=0.003$ after $\tau_{L}$ Myr. For low $\tau_{L}$ = 10 Myr (Row 1, Figure \ref{fig:laplace:tau}) the mantle is warm enough that all of the rheological models are able to push it into its HSE ($T_{m}\approx1800$ K, see Figure \ref{fig:heating_temp}). The state of Io's mantle at the time of initialization of eccentricity falls within the large Maxwell dissipation contours of Figure \ref{fig:shear_visc}. However, if the mantle is allowed to cool for longer (Row 2), then the Maxwell model is not able to produce enough heat to reach HSE. This, coupled with lower dissipation at lower temperatures, leads to a runaway cooling effect that is only countered by the (slowly shrinking) radiogenic heating. Since we consider Io to currently be in a hot state \citep{Morabito1979, Keszthelyi2007a, Spencer2007}: this implies that the Laplace resonance must have initiated shortly after planet formation if Io's mantle has a Maxwell response. If, however, the mantle material is better modeled by an Andrade mechanism, then the Laplace resonance could have initialized much later in Io's cooling (Row 3). \par

\begin{figure*}[hbt!]
    \centering
    \label{fig:laplace:eccen}
    \includegraphics[width=1\linewidth]{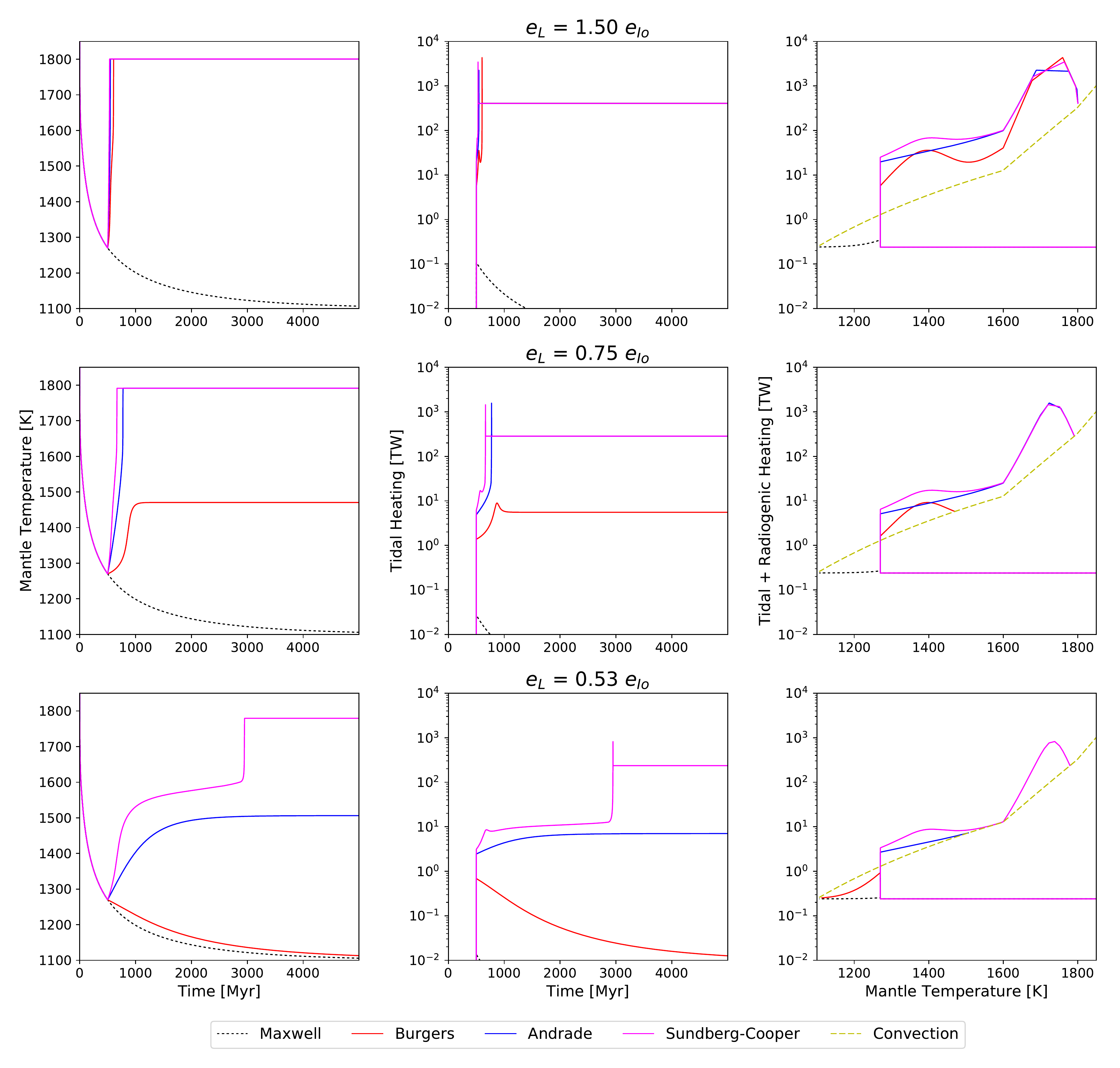}
    \caption{With the same initial state as described in Figure \ref{fig:laplace:tau}, we vary the forced eccentricity that kicks on after $\tau_{L} = 500$ Myr. As the forced eccentricity decreases, some of the rheological models are no longer able to find their HSE. Instead they may find moderate temperature equilibrium. We specifically point out Row 3 where the Sundberg--Cooper model slowly heats as it climbs its quasi-equilibrium described in Section \ref{sec:equilibrium}, while Andrade never escapes this quasi-equilibrium zone in the allotted time.}
\end{figure*}

A similar story can be told if one instead considers the forced eccentricity to be variable at a fixed $\tau_{L}$. Figure \ref{fig:laplace:eccen} shows three different values of forced eccentricity that are allowed to kick on after $\tau_{L} = 500$ Myr. Changing the forced eccentricity has the effect of modifying the difference between the tidal heating and convective cooling curves (see Column 3 in Figure \ref{fig:laplace:eccen}). This difference will affect the location and longevity of various equilibria (recalling that tidal-convective equilibrium points may disappear entirely if tidal forcing drops too low). \par

Overall, the ability of modern rheologies to extend Io's quiescent pre-tidal state implies greater freedom among Laplace resonance formation models. Instead of restricting the assembly of the Laplace resonance to a short time period right after accretion, rheologies like Andrade with enhanced low-temperature dissipation mean that Io could have gone significantly longer without tidal activity and still have achieved the active state seen today. While higher-than-present excursions in tidal forcing also allow longer cooling times at the start, their ability to restore tidal activity is limited, because they must often be of both high intensity and long duration to warm up a cool and unresponsive mantle. Andrade and Sundberg--Cooper mantles recover better in either circumstance: whether the present forcing is the maximum or whether there have been elevated states in the past. Using the same logic, if Io's interior is rather found to be better modeled by a purely Maxwellian rheology, then the Laplace resonance must have initialized within the first 100 Myr after formation.\par

The orbital distance of Io between Jupiter and its neighboring moons is expected to migrate throughout its history. \citet{HussmannSpohn2004} showed that such migrations are possible, along with the previously noted periods of sinusoidal eccentricity variations. This will impact the tidal output within Io and may change the numerical values of the last few paragraphs. Solving for an unknown initial orbital conditions ($a_{0}$, $e_{0}$) can be challenging even in a binary coupled tidal--orbital system with varying internal viscosity. The presence of the Laplace resonance further complicates the obtaining of meaningful solutions for the initial conditions. Therefore, we leave a fully coupled thermal--orbital model with migration for future analysis, but note that the general phenomenological dichotomy between Maxwell and Andrade will still remain in such studies. \par 

Europa and Ganymede are equal partners in the Laplace resonance, and will also benefit from the overall tidal resilience that the Andrade anelasticity component provides. Without severe past forcing episodes, initiating Europa's water ocean from a cold-start scenario can be problematic, because insufficient tidal heating may occur without the added flexibility of a mechanically decoupled shell. In upcoming work, we plan to address how the modifications of Laplace resonance timing may extend beyond Io, out to its neighboring ice--silicate hybrid moons.\par  

Perhaps the most important consequence of this phenomenon, relaxing the time restrictions on when resonance assembly can later lead to tidally active states, is not for Io itself but for exomoons generally. By making it more likely that a diverse range of dynamical capture scenarios and timings lead to meaningful tidal activity in the future, we find that the Andrade and Sundberg--Cooper rheologies can play a significant role in allowing numerous exomoon systems to be tidally warmed across the Galaxy. They initially help prevent bodies from freezing out, and they later help catch moons that do slip temporarily in the direction of such embrittlement. Overall, this may be very good news for maintaining exomoon niches useful for habitability, both on both silicate and ice--silicate hybrid objects.     

\subsection{Frequency Domain} \label{sec:results:frequency}
\begin{figure*}[hbt!]
    \centering
    \label{fig:freq_domain}
    \includegraphics[width=1\linewidth]{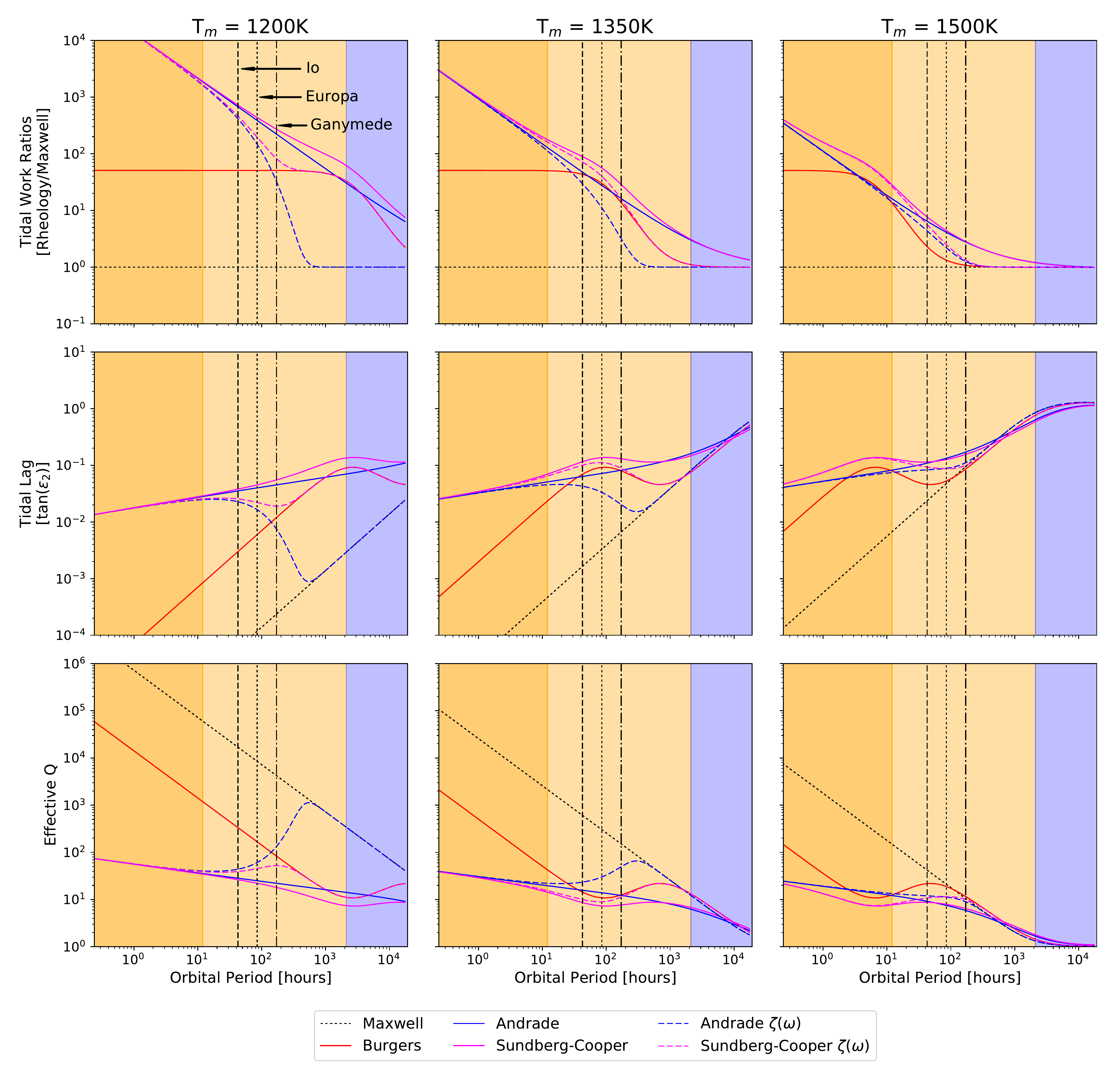}
    \caption{The ratio of a particular rheology's heat production to Maxwell's dissipation is shown (Row 1) at a fixed mantle temperatures of 1200 K (Column 1), 1350 K (Column 2), and 1500 K (Column 3), as are the tangent of the tidal lag $\epsilon_{2}$ (Row 2, see Equation \ref{eq:lag}), and the effective $Q$ values, where $Q^{-1} = \sin{\epsilon_{2}}$ \citep{Efroimsky2012a}. We emphasize different regions of the frequency domain: frequencies that might be obtained in laboratory studies are indicated by dark orange. Moons and short-period exoplanet orbits are designated by light orange. Orbits too far away for strong heliocentric tidal heating are marked in blue. These results were produced assuming an Io-like object orbiting a Jupiter-like host.}
\end{figure*}

Solar system moons like Io have short-period orbits and are considered the most likely massive objects to experience significant tidal forces in our solar system. However, the discovery of short-period exoplanets opens a new area of potentially tidally active worlds. The heliocentric periods of exoplanets have been found to be as short as several hours \citep{Muirhead2012}. \citet{Henning2009} found, on the other hand, that exoplanets may still experience significant tidal activity, in comparison to radionuclide heating, out to periods of $\sim100$ days around typical G- and K-type stars. Before an in-depth study of exoplanets is considered, it is important to ascertain the effect that the rheological models under consideration in this work have in frequency space. In Figure \ref{fig:freq_domain}, we show the tidal dissipation within an Io-like world orbiting a Jupiter-like host over a range of orbital periods. For comparison to other studies we also show the tidal lag produced by the delay between applied shear stress and resultant strain. This lag is sensitive to frequency and exhibits characteristics specific to each rheology \citep[see][]{Efroimsky2012a}. As many dynamicists may be more comfortable working with $Q$ values, we also calculate an effective, not fixed, $Q^{-1}(\omega)=\sin{\epsilon_{2}(\omega)}$. \par

The ratio between tidal heat produced by each non-Maxwell rheology and Maxwell itself is shown to highlight the manner and extent by which models diverge from Maxwell in the high-frequency limit. All other rheologies approach Maxwell in the low-frequency limit, but not before passing outside the band where planetary tides are relevant (outside the light orange shaded region). Within the waveband most relevant for tides, differences from Maxwell are typically of the order of 10$^{2}$--10$^{3}$, and differences amongst the non-Maxwell outputs of the order of 10$^{1}$. Therefore, the choice of rheology can easily overwhelm other errors such as from higher order terms in $e$, global inhomogeneities, or higher order spherical harmonics, each of which often act at the 0.1--2$\times$ level of error. This is particularly important for moons, exomoons, and binaries in the class of trans-Neptunian objects, all of which have the shortest typical periods and thus the greatest rheological choice sensitivity. \par 

Laboratory work finds that the Andrade mechanism's parameters may have their own frequency dependence (see Section \ref{sec:andrade_freq_dependence}). To capture this potential dependence, we also examine both Andrade and Sundberg--Cooper subjected to a frequency-dependent $\zeta(\omega)$, where $\zeta$ is increased exponentially below a critical frequency corresponding to $\sim$1 day$^{-1}$. We emphasize the impact that a frequency-dependent $\zeta$ can lead to, while acknowledging that the full nature of any such $\omega$ dependence will require more analysis than we present here. The Andrade anelasticity can produce strong divergences from the Maxwell and Burgers models at lower periods (higher frequencies). The frequency-dependent $\zeta$ does temper the Andrade response at long periods, but it is precisely because the transition might occur right across the band of Io-like periods that it will be important to determine whether this $\zeta(\omega)$ dependence is real for Io conditions. \par

\subsection{Andrade Parameter Phase Space}\label{sec:results:andrade}
A key challenge for the Andrade model arises from the fact that its two main controlling terms, $\alpha$ and $\zeta$, are not directly associated with classical material property values such as viscosity or shear modulus. They are in some respects equally fundamental, if obscure, material properties, which must be measured in the laboratory to be known, instead of being derived from other properties. This disconnection mirrors the fact they measure the activity of different microphysical events. However, because they are mathematically defined, there remains a gap in being able to link $\alpha$ and $\zeta$ to plain-English meanings, something more easily achieved for viscosity or shear modulus. Describing $\zeta$ as the ratio of the Andrade timescale to the Maxwell timescale does little to help this situation. Prior to this section we used the nominal values of $\alpha=0.2$ and $\zeta=1$. Exploring the behavior of a system when $\alpha$ or $\zeta$ is varied helps move toward understanding these terms, via understanding what they do to outcomes when manipulated. We will explore in future work how the transition from silicate to ice dissipation may perturb these peak dissipation points. \par

\begin{figure*}[hbt!]
    \centering
    \label{fig:andrade_params}
    \includegraphics[width=0.8\linewidth]{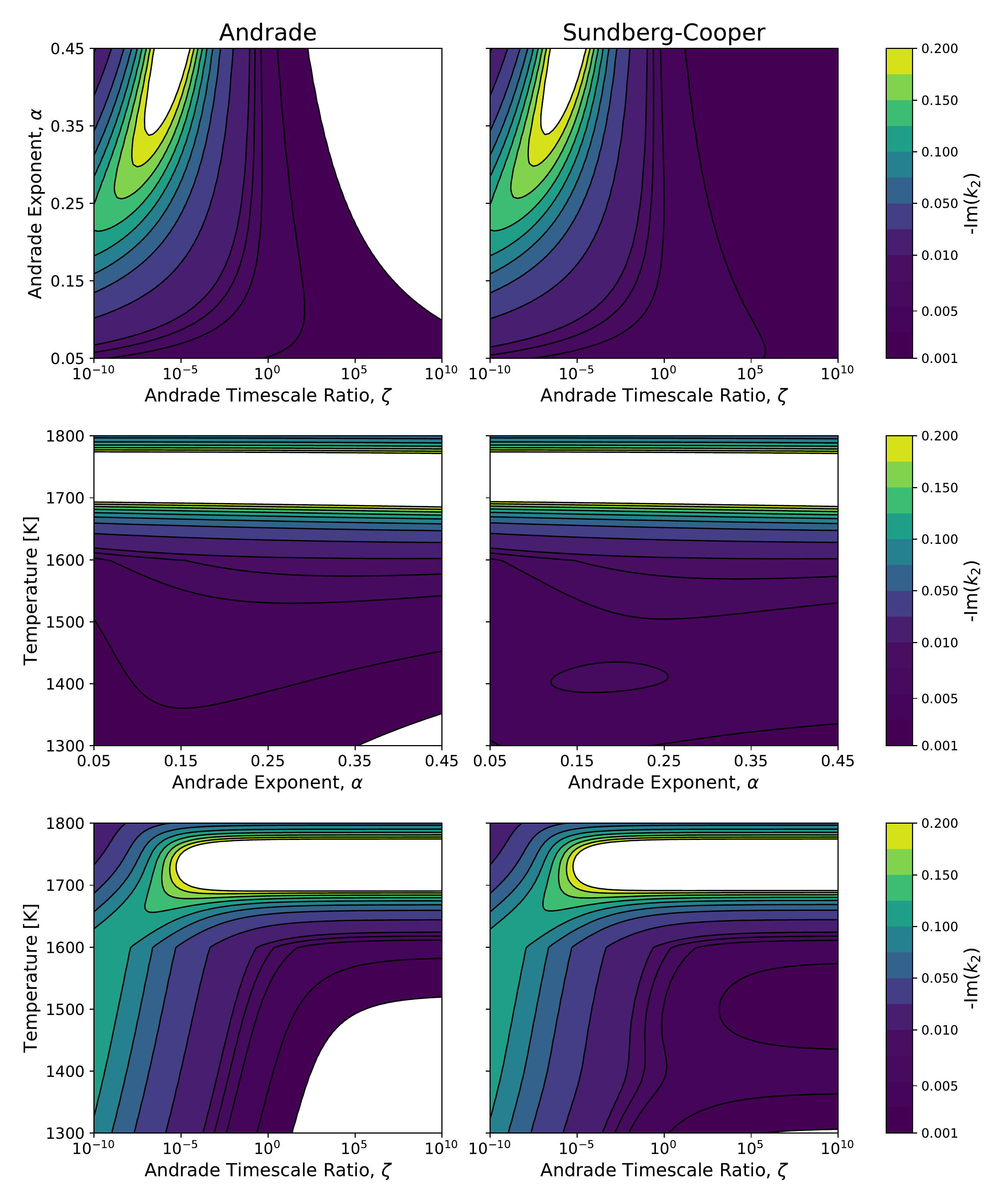}
    \caption{Tidal dissipation, via $-$Im$(k_{2})$, is mapped over the two Andrade empirical parameters (Row 1) as well as temperature (Rows 2 and 3). The Andrade exponent, $\alpha$, is relatively well constrained in the range of 0.1--0.4, based on material composition \citep{Fontaine2005}, whereas the Andrade timescale ratio, $\zeta$, is not, to the authors' knowledge, nearly as constrained. Indeed, the difference between the Andrade and Maxwell timescales will be dependent upon the dominant creep mechanism, which will vary depending upon many circumstances such as pressure, temperature, and stress. For Io, we expect diffusion creep to be dominant, and thereby assume a nominal value of $\zeta \approx 1$ \citep[see discussion in][]{Efroimsky2012a}.  To compensate for this ill-constrained ratio, we show a large domain. Dissipation peaks at high values of $\alpha$ and about a critical value of $\zeta \sim10^{-6}$. In the temperature domain, dissipation is dominated by partial melting for $T > 1600$ K. However, a dampening effect in this region is achieved at low $\zeta$. Rheological effects dominate both models at $T < 1600$ K. A peak in the Sundberg--Cooper model appears at a moderate temperature ($T = 1400$ K) in the range $0.15 < \alpha < 0.25$. This temperature corresponds to the secondary tidal-heating peak seen in Sundberg--Cooper in Figure \ref{fig:heating_temp}.}
\end{figure*}

Being an exponential parameter, $\alpha$ has a greater impact than $\zeta$ upon the rheological response if all else is left constant. $\alpha$ is well constrained between 0.1 and 0.4 (see Section \ref{sec:rheo_response}), but variation within that range can lead to considerable changes. We find there is a narrower range of $\alpha$ that peaks dissipation, but only in certain temperature and/or frequency ranges. The dependence of the Andrade mechanism on temperature and frequency is implicitly affected by $\alpha$ via the term $(\andcomp\andvisc\omega)^{-\alpha}$ in Table \ref{tbl:creep_functions}. Row 2 of Figure \ref{fig:andrade_params} shows the secondary peak of dissipation in the Sundberg--Cooper model at about 1350 K $< T <$ 1450 K and 0.15 $< \alpha <$ 0.25. The island nature of this peak is related to $\voigtcomp$ and $\voigtvisc$ in the Voigt--Kelvin element in Sundberg--Cooper. However, the peak is due to the Andrade mechanism because it can be seen in the same row for the pure Andrade model, centered around $\alpha\approx0.15$. The same temperature range has a swath (going from low to high $\zeta$ values) of moderate dissipation in the $\zeta$ domain (Row 3 of Figure \ref{fig:andrade_params}). The large peak seen in Row 3 of Figure \ref{fig:andrade_params} between 1700 and 1800 K is due to partial melting and is largely independent of rheology. We do note that low values of $\zeta$ dampen this effect ($<10^{-6}$). A much more dramatic dissipation peak is seen about a critical $\zeta$ value of $\sim10^{-6}$ (see Row 1 of Figure \ref{fig:andrade_params}). Interestingly enough, this $\zeta$ value is close to measurements by \citet{Jackson2004} and recently explored in a tidal context by \citet{BiersonNimmo2016}. The strength of this peak is amplified by larger $\alpha$ values. \par

\begin{figure*}[hbt!]
    \centering
    \label{fig:andrade_freq}
    \includegraphics[width=0.9\linewidth]{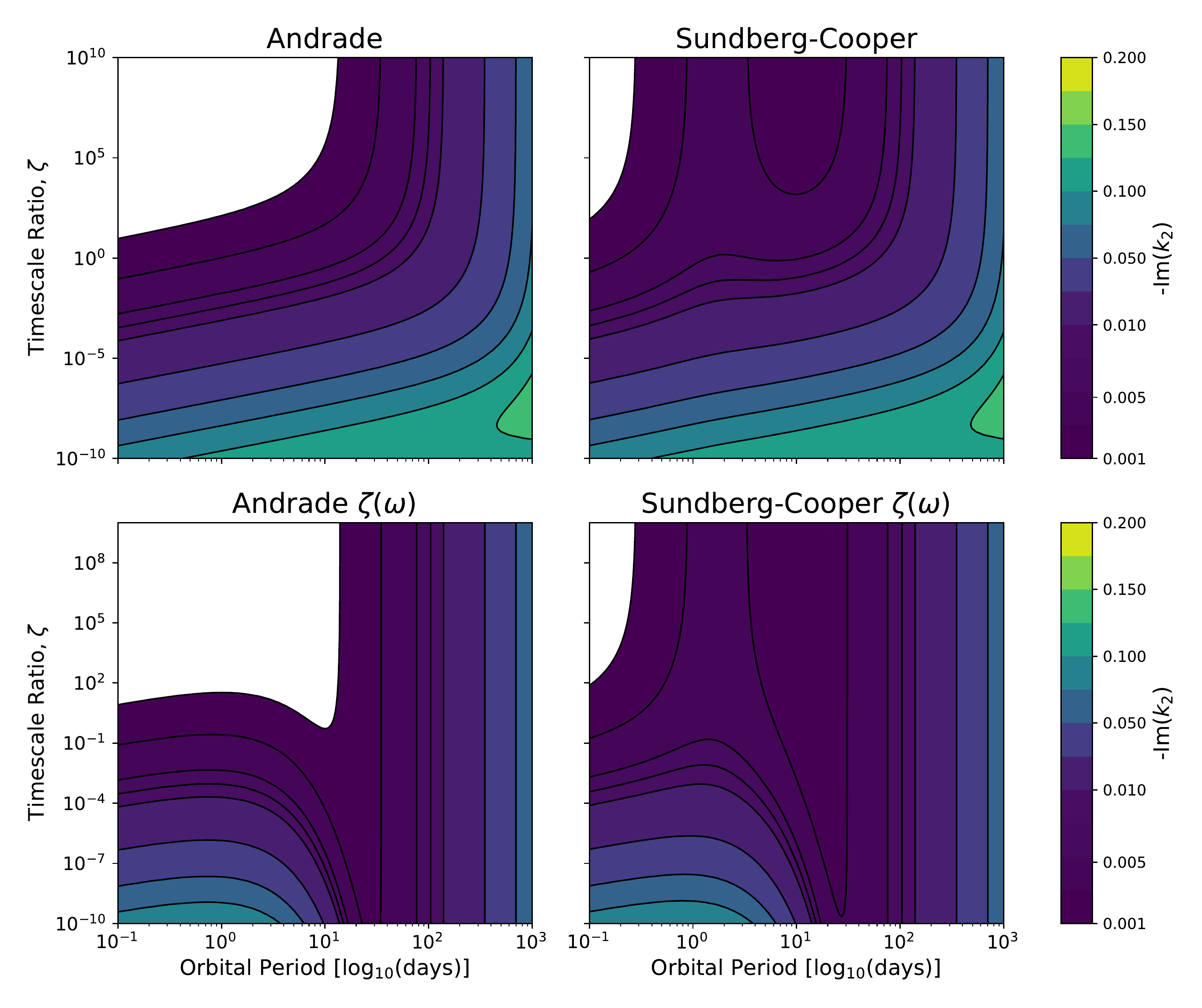}
    \caption{The Andrade timescale, $\zeta$, is varied along with the orbital period. Dissipation is represented by the imaginary component of the Love number. We emphasize the following features. (1) The Sundberg--Cooper model carries moderate dissipation for many values of $\zeta$ and period. (2) The secondary peak from the Burgers component within Sundberg--Cooper produces a moderate dissipation that exists independent of $\zeta$ around a critical period of $\sim$3 days. (3) The frequency-dependent version of the Andrade mechanism will, as expected, lead to no dependence on $\zeta$ below the critical frequency. (4) The selection of a particular $\zeta$ value may lead to relatively consistent dissipation values for drastically different orbital periods. This latter point may help explain the consistent values of $Q$ that are measured for the Moon at its two different tidal frequencies \citep{WilliamsBoggs2008,Efroimsky2012b}. Note that we do not expect dissipation to be strong at large orbital periods (right side of each subplot), because the equation for tidal heating (Equation \ref{eq:tidal_work}) is inversely proportional to several orders of semimajor axis, which will suppress any increase in $-\text{Im}(k_{2})$.}
\end{figure*}

Figure \ref{fig:andrade_freq} shows that the peak about $\zeta\sim10^{-6}$ is mirrored in the orbital-period domain. This peak leads to similar values of $-\text{Im}(k_{2})$ for a large range of orbital periods. However, this consistency is lost if the frequency-dependent Andrade mechanism is utilized. By allowing $\zeta$ to increase below a critical frequency (Row 2 of Figure \ref{fig:andrade_freq}), the Andrade mechanism reduces to the Maxwell viscoelasticity and the $\zeta$ dependence of $-\text{Im}(k_{2})$ is lost. The specific value of this critical frequency (discussed in Section \ref{sec:andrade_freq_dependence}) will be an important consideration. If Figure \ref{fig:andrade_freq} were reproduced with $\omega_{crit}\sim1 \text{yr}^{-1}$ instead of $\omega_{crit}\sim1 \text{day}^{-1}$ the region of frequency independence would be shifted to the right. This would again allow similar dissipation values for many frequencies and may be one explanation as to why we measure similar $Q$ values at frequencies of $\sim1$ month$^{-1}$ and  $\sim1$ yr$^{-1}$ frequencies in our Moon \citep[e.g.,][]{WilliamsBoggs2008}. \par

\subsection{Implications for Exoplanets} \label{sec:results:exoplanets}
Numerous investigations of tidal activity on extrasolar planets have been conducted, with a range of topics from the behavior of gas giants \citep[e.g.,][]{Behounkova2010, Behounkova2011, Remus2012a, Remus2012b, StorchLai2014}, to tidal alterations of system dynamics \citep[e.g.,][]{Lecoanet2009, Matsumura2010, Cebron2011, Bolmont2015, Turbet2017}, to tidal alterations of habitability \citep{Jackson2008, Jackson2008a, Barnes2008, HellerArmstrong2014, Barnes2013, Kopparapu2014}, issues of spin dynamics \citep{Ferraz-Mello2008, Correia2008, Efroimsky2012a, Cunha2015}, and the role of tides on exomoons \citep{Namouni2010, HellerBarnes2013}.  Many such studies naturally begin with frequency-independent internal models, but an increasing number consider viscoelastic models \citep{Henning2009, Behounkova2010, Behounkova2011, Remus2012a, Remus2012b, HenningHurford2014, Auclair-Desrotour2014, ShojiKurita2014, Correia2014, MakarovEfroimsky2014, DriscollBarnes2015, Makarov2015}. Countless more studies rely upon reasonable selections of tidal dissipation terms in order to inform simulations of system dynamics. For solid planetary objects, a detailed study is eventually needed to constrain which rheological models are best under the stress, pressure, and compositional conditions that are applicable to exoplanets and exomoons. Indeed, studies of the Earth tell us that multiple rheological models may be needed as one goes deeper into an exoplanet's interior. Higher pressures will surely change the microphysical mechanisms that govern the rheological response \citep{KaratoSpetzler1990}. We currently must rely mainly on analytical and numerical modeling when exploring the interiors of extrasolar planets, particularly worlds in the super-Earth category not represented in our solar system \citep{Valencia2007}. It is not yet known how well laboratory results on the viscosity of peridotite can extend to high-pressure phases such as post-perovskite \citep{Murakami2004}, which may play a large role in super-Earths.  \par

Increasing data showing planets of terrestrial density around Sun-like stars suggest that there is a large population of exoplanets that may have Earth-analog interiors \citep[e.g.,][]{Morton2016}. More interesting for tides is the growing number of short-period planets that appear to have non-zero eccentricity \citep[e.g.,][]{DawsonFabrycky2010, Rivera2010a, Berta2011, Anglada-Escude2012}. These eccentric, short-period orbits should circularize quickly through tidal dissipation. Severe early scattering may be one explanation \citep{FordRasio2006, FabryckyTremaine2007, Wu2007, Chatterjee2008, Nagasawa2008, Triaud2010, Winn2010, WuLithwick2011, Matsumura2013}. Otherwise, since many of the host stars involved are not young, then these eccentric orbits must: (1) have formed recently, (2) be pumped by nearby companions \citep{Zhang2013}, or (3) have a tidal dissipation that is weaker than expected \citep{HenningHurford2014}, or else (4) the non-zero eccentricities are observational artifacts \citep{ShenTurner2008, Pont2011, Zakamska2011}. The findings of all these works suggest that dissipation mechanisms will be an important component in addressing this puzzle.\par 

Increased tidal dissipation from the Andrade and Sundberg--Cooper rheologies generally acts in opposition to solving questions surrounding eccentric short-period objects. Any increase in tidal dissipation should at first sight translate into an increased fraction of circular orbits. This could be compensated for by more unseen perturbers. However, a less ad hoc amelioration may come from increased dissipation simply translating into more rapid evolution of mantle temperatures into lower-dissipation partial-melt states (such as an emergent magma ocean). Variations in the $Q$ value for the exoplanet's host star will also impact the speed of this evolution. Improved rheologies also allow for long-term equilibrium at moderate tidal heating (see the Sundberg--Cooper/Burgers secondary peak in Figure \ref{fig:heating_temp}.)\par 

\begin{figure*}[hbt!]
    \centering
    \label{fig:exoplanet:k_type}
    \includegraphics[width=0.9\linewidth]{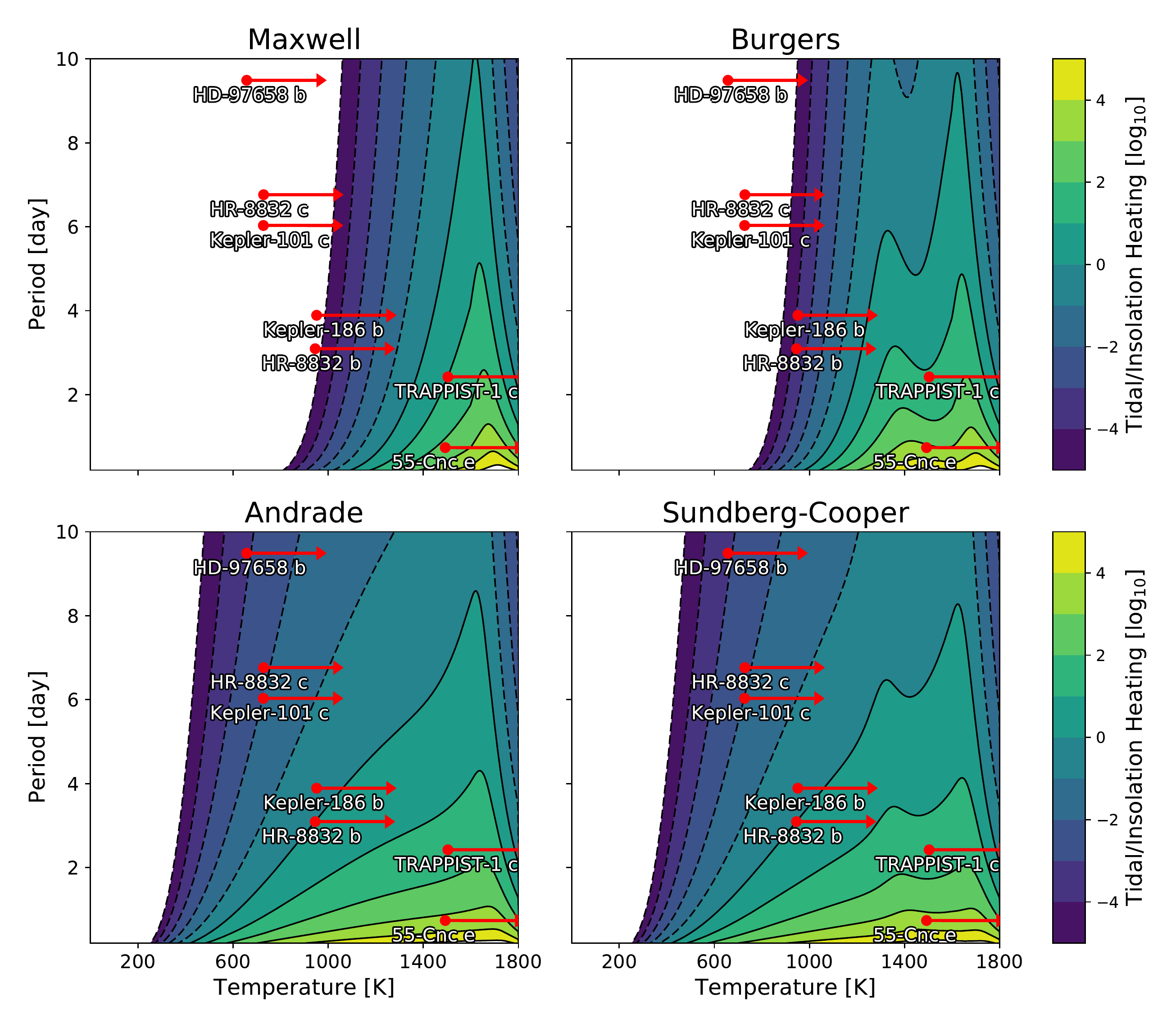}
    \caption{A plot of orbital period vs. mantle temperature with contours of tidal heating over insolation heating. Overlaid on the image are several exoplanets that are plotted with their measured period, and a calculated equilibrium surface temperature (red dots) for a K-type star assuming a planetary albedo of 0.3, and no atmosphere. The arrows represent an increase in temperature from the surface to the mantle where the tidal dissipation is expected to occur. The Earth's mantle temperature increases by thousands of degrees with increasing depth. We can only put a lower constraint on any exoplanet's mantle temperature (red lines). The underlying ratio of tidal heating to insolation is not specific to any of the selected exoplanets, rather it is calculated for a hypothetical rocky planet that has a mass ($M = 3.8$ M$_{E}$ ) and radius ($R = 1.5$ R$_{E}$) equal to the average of the plotted planets. For illustration the planets were chosen based on similar masses and radii, with priority to multi-planet systems where tidal resonances are more likely. We can see that the cooler planets are greatly impacted by an Andrade-like transient mechanism. The relative importance of the rheologies to one another is independent of the eccentricity used.}
\end{figure*}

\begin{figure*}[hbt!]
    \centering\label{fig:exoplanet:m_type}
    \includegraphics[width=0.9\linewidth]{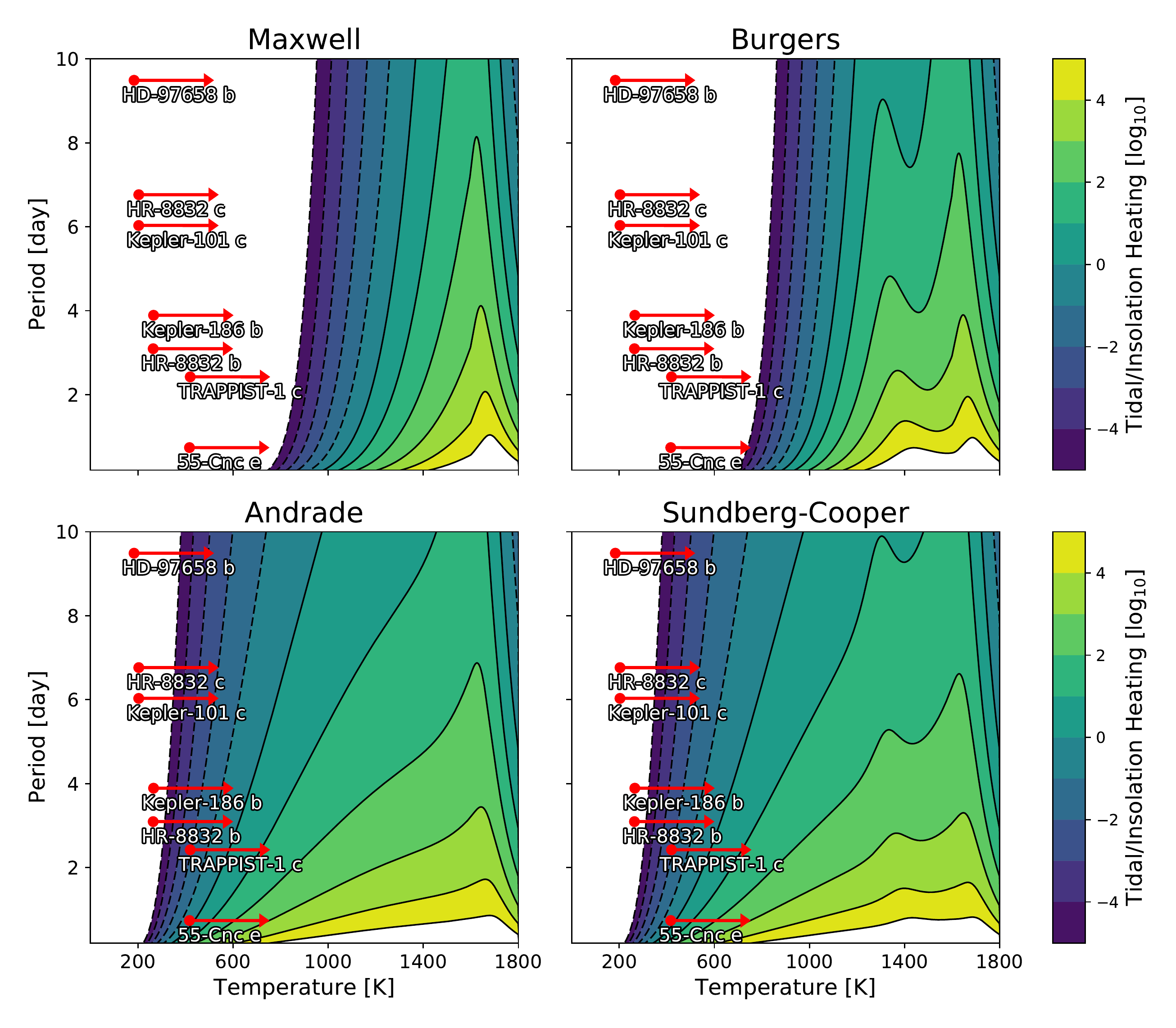}
    \caption{The same methodology that was used in Figure \ref{fig:exoplanet:k_type}, except the star is changed to a main-sequence M-type star. While the much cooler star decreases the surface temperature of any orbiting planets, it will also decrease the magnitude of tidal heating due to the reduced primary mass. The Andrade mechanism is now critical to maintaining large tidal heating in planets that have a mantle temperature comparable to their surface temperature.}
\end{figure*}

If the rheological models explored in this paper are applicable to Earth-mass or larger terrestrial planets, then we can begin to perform order-of-magnitude comparisons. Figures \ref{fig:exoplanet:k_type} (for a K-type star) and \ref{fig:exoplanet:m_type} (for an M-type star) show how tidal heating caused by non-zero eccentricity may overcome insolation heating from a host star for a phase space of orbital period vs. interior temperature. The tidal heating is calculated using Equation \ref{eq:tidal_work}. This formulation assumes that the planet is in a 1:1 spin-orbit resonance. If the planet is in a different spin-orbit resonance (or in between resonances) then there will be additional terms, each with a unique frequency dependence \citep{Ferraz-Mello2008, EfroimskyMakarov2014, Saxena2018}. It is expected that exoplanets may fall into different resonances depending upon their initial orbital state \citep{Rodriguez2012}. Nearby companions could also influence which, if any, resonances a planet may find accessible \citep{Turbet2017}.

To illustrate the possible role of the Andrade and Sundberg--Cooper rheologies, we overlay the location of several currently discovered exoplanets that share roughly similar physical parameters. Surface equilibrium temperatures of exoplanets are shown with a rightward line indicating the uncertainty in the increase in temperature from surface to interior. For Earth, the temperature jump between the surface and upper mantle is roughly 1000 K, with a shallow adiabatic gradient thereafter. For exoplanets this will depend on the internal heat flux, the lithosphere structure, and the possible existence of heat-pipe behavior. At moderate mantle temperatures and for short-period orbits, the tidal heating will be strong no matter which rheological model is used. For longer periods and/or cooler planets, the rheological differences become a key factor that should be considered in future studies. The shaded contours in Figures \ref{fig:exoplanet:k_type} and \ref{fig:exoplanet:m_type} are chosen specifically as case-independent ratios of tidal heating to insolation. One may compare between the two figures the degree to which various objects are enveloped by contoured regions. A general trend toward increased tidal heating using realistic rheologies is evident, and is particularly significant for cooler stars.  \par

In Section \ref{sec:results:str_visc} we demonstrated how $M_{sec}$ acts as a control on the extent to which varying rheology features are expressed during thermal evolution. Objects with $M_{sec} \sim100$ M$_{Io}$ typically have the greatest expression of Andrade-mechanism dissipation, while objects with $M_{sec} >$ 10 M$_{E}$ express only the shoulder of the Andrade mechanism band. This is true regardless of forcing frequency or host mass. Despite this, even expressing part of the Andrade-mechanism dissipation will lead to greater tidal resilience for exoplanets, or especially exomoons, when utilizing a model containing the Andrade anelasticity. \par

 But this mass dependence does mean that for silicate exomoons the lessons we take from modeling Io may be extensible rather broadly. We therefore predict that use of modern material models will increase the number of exomoons that can endure in tidally active states in the broader Galaxy, across a wide host of orbital histories. The notion that volcanic activity is more common via this update in material modeling is an attractive and potentially observable concept. Likewise, tidally induced water oceans also expand in resilience, because Andrade has been found to apply to ice just as it does to silicate. The same principle of response broadening upon tidal--orbital interactions also applies, and will be studied in detail for ice worlds in our future work. \par
 
 The specific magnitudes of tidal heating presented in this section may change when compressibility is considered. However, the overall shape of the response in the temperature and frequency domain will be largely retained. The importance of one rheological model over another will be just as valid when a more robust exoplanet interior is considered. The main idea demonstrated here is that application of the Andrade and Sundberg--Cooper rheologies cause more exoplanets to be tidally active than a Maxwell application, largely regardless of other inputs. \par
 
 \subsection{Radiogenic-mediated Equilibrium Loss}\label{sec:results:radio_loss}
 
\begin{figure}[hbt!]
    \centering
    \label{fig:exoplanet_time}
    \includegraphics[width=1\linewidth]{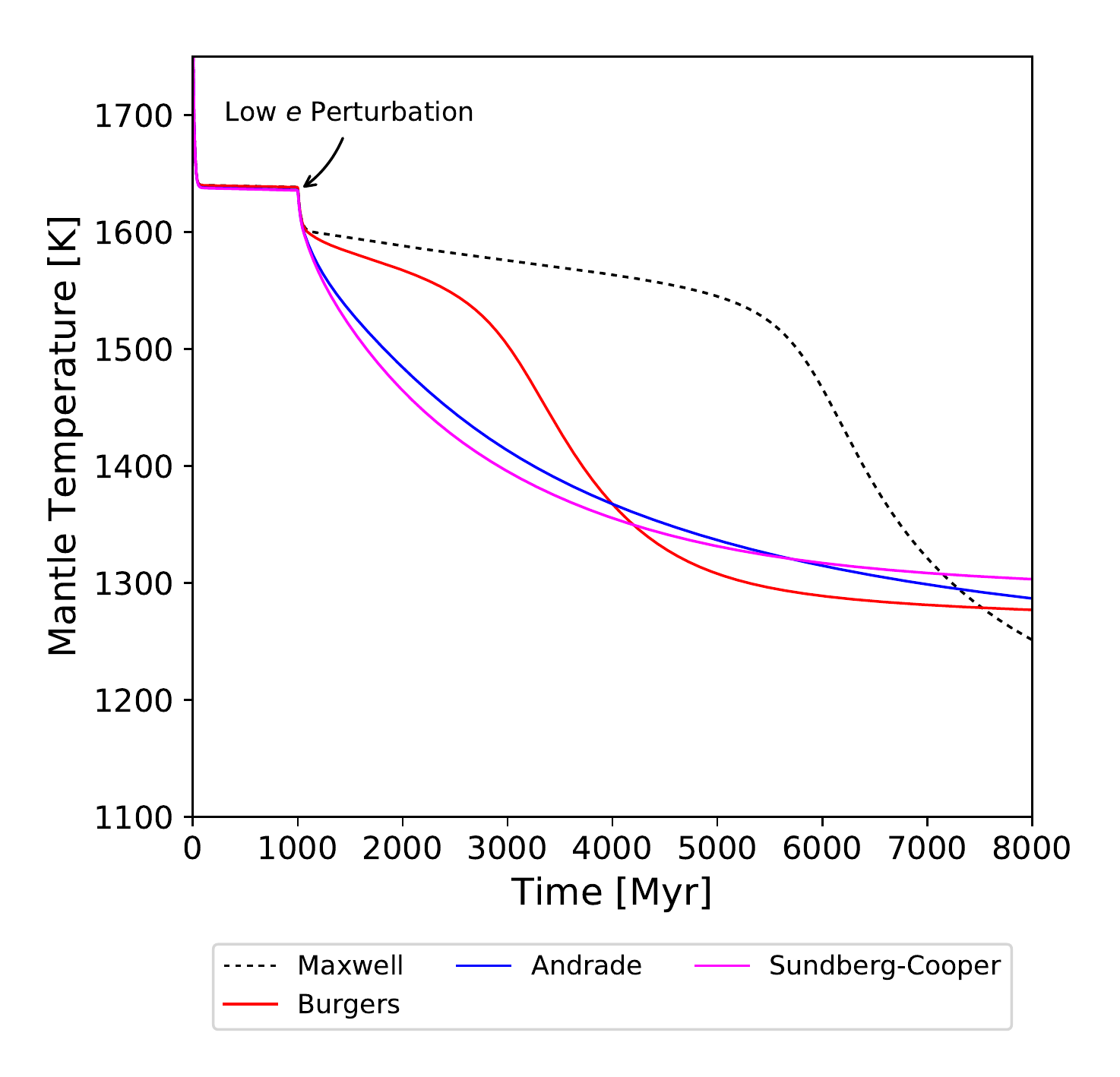}
    \caption{To illustrate a concept we refer to as radiogenic-mediated equilibrium loss, a super-Earth planet ($M$ = 3.80 M$_{E}$, $R$ = 1.45 R$_{E}$, and $a = 0.1$ au) is initially subjected to $e$ = 0.20 around an M dwarf host. At 1,000 Myr, $e$ is reduced to 0.07. Radiogenic heating decays exponentially throughout the model (with values matching the lower mantle of modern Earth in at 4500 Myr). At a time that could be many millions of years after the initial loss of eccentricity, the planet's total heating (tidal and radiogenic) drops below the convective curve (see Figure \ref{fig:heating_temp}). Depending on a rheology's low-temperature response, this can cause rapid, but delayed, cooling. In this figure the Maxwell rheology, due to a higher peak, is able to withstand for the longest time after the reduction of eccentricity. But it also drops faster than the other models.}
\end{figure}

On looking at Figure \ref{fig:heating_temp} one will notice a small difference between the largest peaks of the Maxwell and Andrade models. Due to the log-scaling, this difference turns out to be $\sim100$ TW. Generally this does not influence the thermal history, since any time evolution will quickly progress through this region on the way to either an HSE state or secular cooling. Our simulations show that an evolutionary model may only stay near the peak for a handful of 100,000 yr time steps (as seen in the very jagged features around this region in Row 1, Column 3 of Figure \ref{fig:laplace:eccen}). However, an interesting phenomenon can occur when one considers a planet that is already at the HSE and induced to lose most of its forced eccentricity (much the same as Row 1 of Figure \ref{fig:step_func}). The reader can imagine the impact of this scenario by slowly shifting the tidal heating curves in Figure \ref{fig:heating_temp} down, while keeping the convection curve constant. There is a certain critical eccentricity for each rheology (dependent upon the system's parameters) where the convection curve just barely grazes the top of each peak. A small perturbation will send the planet into secular cooling. There are countless orbital scenarios that could cause such a perturbation. Instead we point to a purely internal one: the slow decay of radiogenics. If a planet is equilibrated at the HSE above this critical eccentricity, and then suddenly loses much of its eccentricity (perhaps due to the ejection of the perturber that was pumping it) then its tidal heating may fall to the point where it is on the verge of passing through this region. The interior will continue to lose heating from the loss of isotope concentrations over time. This can cause a planet to eventually pass through the critical point, leading to a loss of its hot state. This could occur many millions of years after the actual orbital event that triggered the inevitable outcome. Since the Maxwell model has the largest peak heating value, it will be the last to succumb. But, since the Maxwell model has very weak heating at cooler temperatures it will also have the most dramatic loss of heating. We explore this phenomenon by considering a super-Earth exoplanet ($M$ = 3.80 M$_{E}$, $R$ = 1.45 R$_{E}$, and $a$ = 0.1 au) orbiting an M-type star with an initial $e$ = 0.20. After 1000 Myr we reduce eccentricity to 0.07. After this point there are no other actions imposed on the planet except the convective cooling of the mantle and the decay of isotopes. The aforementioned post-perturbation temperature losses can be seen in Figure \ref{fig:exoplanet_time}. Note that this phenomenon of radiogenic-mediated equilibrium loss is primarily an exoplanet concern, more so than for Io or exomoons, simply due to the larger supply of, and temporal change in, radiogenic elements.\par

Lastly, given the potential for plate tectonics on Earth-analog exoplanets, a unique new aspect of the property of tidal resilience of Andrade-like models arises. One non-orbital form of a transient low-forcing excursion for a planet would be a mantle overturn event, or else the foundering of a major lithospheric cold slab. Such events could induce large-scale transient cooling of a mantle, akin to other orbital perturbations caused by low tidal forcing. The Andrade and Sundberg--Cooper models would greatly help a planet to restore status-quo tidal heating long after such an event, just as they do for other perturbations. \par

\section{Conclusions} \label{sec:conclusion}

Laboratory studies suggest that the simple fixed-$Q$ model, and even the Maxwell viscoelastic model, do not capture many of the intricacies seen in the deformation of real materials. Since these have been the traditional models used in tidal studies, it is necessary to understand the implications that new rheological models have in tidal--orbital modeling. We show that the Andrade anelasticity, which is contained within both the Andrade and Sundberg--Cooper rheology models, is able to generate much larger dissipation at lower mantle temperatures. This increased dissipation can greatly affect the long-term evolution of planets that are experiencing secular cooling before tidal forces are activated. For instance, if an Andrade-like rheology is applicable to the interior of Io, then the 4:2:1 Laplace resonance among the Galilean moons could have assembled much later after Io's formation than allowed by an Io driven solely by the Maxwell model. Any Maxwell response that does not initiate within $\sim100$ Myr after Io's formation will not allow Io to return to the hot state we see today, unless the forced eccentricity was once much greater than the values we measure today. Increased dissipation at lower temperatures also impacts the speed at which Io is able to convert orbital energy to internal heat. While this has the potential to alter the long-term stability of the Laplace resonance, we leave this question for future thermal--orbital coupled work. \par

Prior debates regarding the Laplace resonance, where a fixed-$Q$ for Io has been invoked, miss the likely reality that the $Q$ of Io prior to the onset of strong tides can be vastly higher than $Q$ once tides are active, all for reasons of mantle temperature. This is somewhat analogous to classical friction, whereby it would be erroneous to assume a box's coefficient of sliding friction to be the same as its coefficient of static friction: plausibly starting tides on Io (or handling start--stop scenarios) requires overcoming special geophysical initial conditions. Late-assembly models of the Laplace resonance have been in jeopardy of allowing Io to become too cold to initiate tidal activity, but our results restore the permissibility of these models against this concern. \par

Exomoons, as well as short-period exoplanets, made of similar material to the Earth should also have an Andrade-like response in some or all of their layers. Such increased dissipation may cause tidal heating to become the dominant heat source within exomoon and exoplanet interiors for a larger subset of worlds than previously expected. If long-term eccentricities are occurring for short-period exoplanets, as evidence suggests, then the increased dissipation implied by the material models here presents a mild complication. Increased dissipation would typically imply faster circularization. However, one path to resolution of this issue is that increased dissipation actually translates into planets evolving more rapidly into a low-dissipation partial-melt state.  \par

We find that use of the Andrade and Sundberg--Cooper rheologies leads to enhancement of a property we term tidal resilience, or the ability of ongoing tidal activity to endure for long durations in the face of perturbations. Because the Andrade and Sundberg--Cooper models (as well as the Burgers model to some extent) lead to greater dissipation at low temperatures, they have improved capability for a tidally active interior to recover after a low-eccentricity excursion, or a low-tidal-forcing excursion of any other form. Both having relaxed conditions for timings of resonance assembly that can achieve future tidal activity, as well as overall tidal resilience, are beneficial for maintaining tidal warmth on exomoons, where habitable conditions are often determined by tides, not insolation. \par

The Andrade exponent $\alpha$ leads to the greatest overall changes in both the Andrade and Sundberg--Cooper models, independent of any other considerations. However, if frequency-dependent Andrade parameters are considered, there is a critical timescale ($\zeta\sim10^{-6}$ for Io) that can greatly change dissipation. Rheological dependence on temperature/melt-fraction (indirectly through viscosity and compliance) and frequency (directly) are influenced by both empirical parameters. Temperature couples more strongly with $\zeta$ rather than $\alpha$, leading to larger changes in dissipation. Below a critical frequency, a transformation from the Andrade anelasticity into Maxwell is expected. While the critical frequency in this work leads to significant impact on Io, such a frequency is not excluded from being much lower (months or years). If ever determined, a low critical frequency would force a non-Maxwell state on short-period exoplanets/exomoons. In this same scenario, the seemingly frequency-independent $Q$ of our Moon could be explained by a critical $\zeta$ value if its interior is well modeled by the Andrade anelasticity.\par

It remains true, as always, that further laboratory experiments are the cornerstone on which tidal modeling will continue to improve. If laboratory work continues to point to Andrade-like models for the wide range of materials and temperature--pressure conditions as found to date, we expect this model will grow in application. Similarly, broad application of the Sundberg--Cooper model is most dependent on growing support from laboratory results, which in turn hinges upon continued support for research on mantle-relevant materials. Likewise, continued observations of the heat flow leaving tidally active worlds, such as Io, will allow us to better constrain interior states.\par 

Overall we recommend that the Andrade and Sundberg--Cooper rheologies be strongly considered for any solid-body tidal application when errors finer than 10$\times$ are desired in mapping outcomes back to interior conditions. This is particularly true for masses of 1M$_{Io}$ -- 10M$_E$, mantle temperatures from 1000--1600 K, and across \textit{all} tidally relevant forcing periods.\par

\acknowledgments
We would like to thank Michael Efroimsky, Valeri Makarov, Terry Hurford, Robert Tyler, Christopher Hamilton, Joe Weingartner, and Yuri Mishin for their insights, comments, and suggestions. This work was supported by the George Mason University Physics \& Astronomy Scholarship, the John Mather Nobel Scholar Fund, the NASA Exoplanets Research Program grant NNX15AE20G, and NASA Outer Planets Research Program grant NNX14AR42G.

\clearpage
\appendix

\renewcommand{\arraystretch}{.9} 
\begin{table}[htbp]
    \centering
    \section{Parameters \& Formulae}
    \caption{Key parameters, formulae, and nominal values used in our model.\label{tbl:parameters}}
    \begin{tabular}{llll}
    \hline
    \hline
    \textbf{Symbol}            & \textbf{Value/Formulae}     & \textbf{Unit}              & \textbf{Definition} \\
    \hline
    $G$                        & $6.674\times10^{-11}$      & m$^{3}$ kg$^{-1}$ s$^{-2}$ & Newton's Gravitational Constant \\
    $\sigma_{B}$               & $5.67\times10^{-8}$        & W m$^{-2}$ K$^{-4}$        & Stefan-Boltzmann Constant \\
    $\beta$                    & $\andcomp\left(\zeta{}\andcomp\andvisc\right)^{-\alpha}$ & Pa$^{-1}$ s$^{-\alpha}$  & Andrade Emperical Coefficient \\
    $S$                        & $\alpha!\sin{\left(\alpha\pi/2\right)}$ & Unitless                   & Andrade Constant 1 \\
    $C$                        & $\alpha!\cos{\left(\alpha\pi/2\right)}$ & Unitless                   & Andrade Constant 2 \\
    $\alpha$                   & 0.2              & Unitless                   & Andrade Emperical Exponent (nominal value) \\
    $\zeta$                    & 1                & Unitless                   & Andrade Emperical Timescale (nominal value) \\
    $\lambda$                   & $\left(\maxcomp\voigtvisc\omega\right)^{2} + \left(\maxcomp/\voigtcomp\right)^{2}$ & Unitless & Burgers Parameter \\
    Ra$_{c}$                   & $1100.0$                   & Unitless                   & Critical Rayleigh Number \\
    $\gamma$                   & 0.011                      & Unitless                   & Mantle Viscosity Ratio \\
    $\kappa$                   & $9.16\times10^{-7}$        & m$^{2}$ s$^{-1}$           & Thermal Diffusivity \\
    $\alpha_{V}$               & $3\times10^{-5}$           & K$^{-1}$                   & Mantle Thermal Expansion \\
    $a_{c}$                    & 0.1                        & Unitless                   & Convection Thickness Proportionality \\
    $\omega_{crit}$            & $1\times10^{-4}$           & rad s$^{-1}$               & Critical Andrade Frequency \\
    $\omega=n$                   & $4.11\times10^{-5}$        & rad s$^{-1}$               & Forcing Frequency (assumed to be mean motion) \\
    $Q_{CMB}$                  & see Equation \ref{eq:q_cmb} & W                         & Core--to--Mantle Heating \\
    $Q_{Conv}$                 & see Equation \ref{eq:q_conv}& W                         & Mantle--to--Surface Heating \\
    $M_{m}$; $M_{c}$  & $0.8M_{Io}$; $0.2M_{Io}$                & kg                         & Mantle; Core Mass \\
    $\maxcomp$            & $1.66\times10^{-11}$       & Pa$^{-1}$                   & Unrelaxed Compliance \\
    $\voigtcomp$               & $0.2\maxcomp$              & Pa$^{-1}$                   & Compliance Defect\footnote{The compliance defect is defined such that the relaxed compliance (at infinite time after load) is $J_R = \maxcomp + \voigtcomp$.} \\
    $\tilde{\mu}$              & $19\maxcomp^{-1}/2\rho{}gR_{sec}$ & Unitless                   & Effective Rigidity \\
    $M$                        & $J^{-1}$                   & Pa                         & Rigidity \\
    $\maxvisc$         & $1\times10^{22}$           & Pa s                       & Maxwell Viscosity \\
    $\voigtvisc$                 & $0.02\maxvisc$             & Pa s                       & Voigt--Kelvin Viscosity \\
    $R_{sec}$                  & $1.82\times10^{6}$         & m                          & Io's Radius \\
    $g$                        & 1.79                       & m s$^{-2}$                 & Gravitational Surface Acceleration \\
    $e_{present}$              & 0.0041                     & Unitless                   & Io's Present-day Eccentricity \\
    $a$                        & $4.22\times10^{8}$         & m                          & Io's semimajor Axis \\
    $a_{*}$                    & $7.79\times10^{11}$        & m                          & Jupiter's semimajor Axis \\
    $A$                        & 0.63                       & Unitless                   & Albedo \\
    $\epsilon_{v}$             & 0.9                        & Unitless                   & Emissivity \\
    $L_{*}$                    & $3.85\times10^{26}$        & W                          & Luminosity \\
    $M_{sec}$                  & $8.93\times10^{22}$        & kg                         & Io's Mass \\
    $M_{pri}$                  & $1.9\times10^{27}$         & kg                         & Jupiter's Mass \\
    $T_{sol}$; $T_{br}$; $T_{liq}$                  & 1600; 1800; 2000                       & K                          & Temperatures of Solidus; Breakdown; Liquidus \\
    $k_{m}$                    & 3.75                       & W m$^{-1}$ K$^{-1}$        & Mantle's Thermal Conductivity \\
    $c_{c}$; $c_{m}$      & 444.0; 1260.0                  & J K$^{-1}$ kg$^{-1}$       & Core; Mantle Specific Heat \\
    $T_{m}$; $T_{c}$                     & see Eqns. \ref{eq:mantle_temp} \& \ref{eq:core_temp} & K                    & Temperatures of Mantle; Core \\
    $D_{m}$                    & $8.8\times10^{5}$          & m                          & Mantle Thickness \\
    $f_{tvf}$                  & $85\%$                     & m$^{3}$ m$^{-3}$           & Mantle's Tidal Volume Fraction \\
    \hline
    \hline
    \end{tabular}%
  \label{tab:addlabel}%
\end{table}%

\renewcommand{\arraystretch}{1} 

\begin{table}
\section{Compliance Functions}
\caption{The compliance, or `creep-response', functions (which we denote by $\bar{J}(t)$) for the various rheologies under consideration are shown here in the frequency domain.\label{tbl:creep_functions}}
\begin{center}
\begin{tabular}{ c c }
    \hline
    \hline
	\textbf{Rheology} & \textbf{Creep Function} \\
	\hline
	& \\
    Maxwell & $\displaystyle\maxcomp - \frac{\displaystyle{}i}{\displaystyle\maxvisc\omega}$ \\[20pt]
    Burgers & $\displaystyle\maxcomp - \frac{\displaystyle{}i}{\displaystyle\maxvisc\omega} + \frac{i\voigtcomp}{i - \voigtcomp\voigtvisc\omega}$ \\[20pt]
    Andrade & $\displaystyle\maxcomp - \frac{\displaystyle{}i}{\displaystyle\maxvisc\omega} + \andcomp(i\andcomp\andvisc\zeta\omega)^{-\alpha}\alpha!$ \\[20pt]
    Sundberg--Cooper & $\displaystyle\maxcomp - \frac{\displaystyle{}i}{\displaystyle\maxvisc\omega} + \frac{\displaystyle{}i\voigtcomp}{\displaystyle{}i - \voigtcomp\voigtvisc\omega} + \andcomp(i\andcomp\andvisc\zeta\omega)^{-\alpha}\alpha!$ \\[20pt]
    \hline
    \hline
\end{tabular}
\end{center}
\end{table}

\begin{sidewaystable}
    \section{Complex Love Number}
    \centering
    \caption{We present the expanded imaginary portion of the complex Love number $-\text{Im}(k_{2})$. The presentation of the formulae was designed so that the reader may see how specific components evolve from the Maxwell model to the Sundberg--Cooper model. Depending on the situation, assumptions may be made to eliminate or simplify terms, see the discussion in Section \ref{sec:rheo_response} for more details.\label{tbl:love_numbers}}
   \begin{tabular}{l c}
    \hline
    \hline
	\textbf{Rheology} & \textbf{Love Number, $-\text{Im}(k_{2})$} \\
	\hline
	& \\
    Maxwell & $\displaystyle\frac{\displaystyle3}{\displaystyle2}\frac{\displaystyle\maxcomp\maxvisc\omega\tilde{\mu}}{\displaystyle1 + \left(\maxcomp\maxvisc\omega\right)^{2}\left(\tilde{\mu} + 1\right)^{2}}$ \\[25pt]
    Burgers & $\displaystyle\frac{\displaystyle3}{\displaystyle2}\frac{\displaystyle\maxcomp\maxvisc\omega\tilde{\mu}\left[\left(\voigtcomp\maxvisc\omega\right)^{-2} + \left(\frac{\displaystyle\voigtvisc}{\displaystyle\maxvisc}\right)^{2} + \frac{\displaystyle\voigtvisc}{\displaystyle\maxvisc}\right]}{\displaystyle\left(\voigtcomp\maxvisc\omega\right)^{-2} + \lambda\left(\tilde{\mu} + 1\right)^{2} + 2\frac{\displaystyle\maxcomp}{\displaystyle\voigtcomp}\left(\tilde{\mu} + 1\right) + \left(\frac{\displaystyle\voigtvisc}{\displaystyle\maxvisc} + 1\right)^{2}}$ \\[25pt]
    Andrade & $\displaystyle\frac{\displaystyle3}{\displaystyle2}\frac{\displaystyle\maxcomp\maxvisc\omega\tilde{\mu}\left[1 + \left(\maxcomp\maxvisc\omega\right)^{1-\alpha}\zeta^{-\alpha}S\right]}{\displaystyle1 + \left(\maxcomp\maxvisc\omega\right)^{2}\left(\tilde{\mu} + 1\right)^{2} + \left(\maxcomp\maxvisc\omega\right)^{2(1-\alpha)}\zeta^{-2\alpha}\left(a!\right)^{2} + 2\left(\maxcomp\maxvisc\omega\right)^{2-\alpha}\zeta^{-\alpha}\left[\left(\maxcomp\maxvisc\omega\right)^{-1}S + \left(\tilde{\mu} + 1\right)C\right]}$ \\[25pt]
    Sundberg--Cooper & $\displaystyle\frac{\displaystyle3}{\displaystyle2}\frac{\displaystyle\maxcomp\maxvisc\omega\tilde{\mu}\left[\left(\voigtcomp\maxvisc\omega\right)^{-2} + \left(\frac{\displaystyle\voigtvisc}{\displaystyle\maxvisc}\right)^{2} + \frac{\displaystyle\voigtvisc}{\displaystyle\maxvisc} + \lambda\left(\maxcomp\maxvisc\omega\right)^{-(1+\alpha)}\zeta^{-\alpha}S\right]}{\displaystyle\left(\voigtcomp\maxvisc\omega\right)^{-2} + \lambda\left(\tilde{\mu} + 1\right)^{2} + 2\frac{\displaystyle\maxcomp}{\displaystyle\voigtcomp}\left(\tilde{\mu} + 1\right) + \left(\frac{\displaystyle\voigtvisc}{\displaystyle\maxvisc} + 1\right)^{2} +  \lambda\left(\maxcomp\maxvisc\omega\zeta\right)^{-2\alpha}\left(a!\right)^{2} + 2\left(\maxcomp\maxvisc\omega\zeta\right)^{-\alpha}\left[\left\{\maxcomp\voigtvisc\omega\left(1 + \voigtvisc/\maxvisc\right) + \maxcomp/\voigtcomp\left(\voigtcomp\maxvisc\omega\right)^{-1}\right\}S + \left\{\lambda\left(\tilde{\mu} + 1\right) + \maxcomp/\voigtcomp\right\}C\right]}$ \\[25pt]
    \hline
    \hline
    \end{tabular}
\end{sidewaystable}

\begin{table*}
\section{Complex Rigidity}
\centering
\caption{Complex rigidity functions, derived from $\bar{M} = \bar{J}^{-1}$ using the complex compliance functions ($\bar{J}$(t), see Table \ref{tbl:creep_functions}). Here $\bar{M}$ = $M_1$ + $iM_2$, with $M_1$ = $N_1$/$D^{*}$, while $M_2$ = $N_2$/$D^{*}$. Common denominators $D^{*}$ can be found in Table \ref{tbl:rigidity_function:table_b}. The presentation of the formulae here are designed to mimic that of the complex Love number ($-\text{Im}(k_{2})$, see Table \ref{tbl:love_numbers}). \label{tbl:rigidity_function:table_a}}
\begin{tabular}{ c  c  c }
    \hline
    \hline
	\textbf{Rheology} & \textbf{$N_1$} &  \textbf{$N_2$} \\
	\hline
	& & \\
    Maxwell & $\displaystyle\maxcomp\maxvisc^{2}\omega^{2}$ & $\displaystyle\maxvisc\omega$ \\[20pt]
    Burgers & $\displaystyle\left(\frac{1}{\maxcomp}\right)\left(\lambda + \frac{\maxcomp}{\voigtcomp}\right)$ &  $\displaystyle\maxvisc\omega\left(\frac{\voigtvisc}{\maxvisc} + \left(\frac{\voigtvisc}{\maxvisc}\right)^{2} + \left(\maxvisc\omega\voigtcomp\right)^{-2}\right)$ \\[20pt]
    Andrade & $\displaystyle\maxcomp\maxvisc^{2}\omega^{2}\left(\left(\maxcomp\maxvisc\omega\zeta\right)^{-\alpha}C+1\right)$ & $ \displaystyle\maxvisc\omega\left(\left(\maxcomp\maxvisc\omega\right)^{1-\alpha}\zeta^{-\alpha}S+1\right)$ \\[20pt]
    Sundberg--Cooper & $\displaystyle\left(\frac{1}{\maxcomp}\right)\left[\left(\left(\maxcomp\maxvisc\omega\zeta\right)^{-\alpha}C+1\right)\lambda + \frac{\maxcomp}{\voigtcomp}\right]$ & $\displaystyle\maxvisc\omega\left[\left(\maxcomp\maxvisc\omega\right)^{-(1+\alpha)}\zeta^{-\alpha}S\lambda + \frac{\voigtvisc}{\maxvisc} + \left(\frac{\voigtvisc}{\maxvisc}\right)^{2} + \left(\maxvisc\omega\voigtcomp\right)^{-2}\right]$ \\[20pt]
    \hline
    \hline
\end{tabular}
\end{table*}

\begin{sidewaystable}
    \centering
    \caption{The denominators for the real and imaginary parts of the complex rigidity functions.\label{tbl:rigidity_function:table_b}}
   \begin{tabular}{l c}
    \hline
    \hline
	\textbf{Rheology} & \textbf{D*} \\
	\hline
	& \\
    Maxwell & $\displaystyle1 + \maxcomp^{2}\maxvisc^{2}\omega^{2}$ \\[20pt]
    Burgers & $\displaystyle\lambda + \left(\voigtcomp\maxvisc\omega\right)^{-2} + 2\frac{\maxcomp}{\voigtcomp} + \left(\frac{\voigtvisc}{\maxvisc}+1\right)^{2}$\\[20pt]
    Andrade & $ \displaystyle1 + \left(\maxcomp\maxvisc\omega\right)^{2} + \left(\maxcomp\maxvisc\omega\right)^{2(1-\alpha)}\zeta^{-2\alpha}\left(a!\right)^{2} + 2\left(\maxcomp\maxvisc\omega\right)^{2-\alpha}\zeta^{-\alpha}\left[\left(\maxcomp\maxvisc\omega\right)^{-1}S + C\right]$ \\[20pt]
    Sundberg--Cooper &
    $\displaystyle\left(\voigtcomp\maxvisc\omega\right)^{-2} + \lambda + 2\frac{\displaystyle\maxcomp}{\displaystyle\voigtcomp} + \left(\frac{\displaystyle\voigtvisc}{\displaystyle\maxvisc} + 1\right)^{2} +  \lambda\left(\maxcomp\maxvisc\omega\zeta\right)^{-2\alpha}\left(a!\right)^{2} + 2\left(\maxcomp\maxvisc\omega\zeta\right)^{-\alpha}\left[\left\{\maxcomp\voigtvisc\omega\left(1 + \frac{\voigtvisc}{\maxvisc}\right) + \frac{\maxcomp}{\voigtcomp}\left(\voigtcomp\maxvisc\omega\right)^{-1}\right\}S + \left\{\lambda + \frac{\maxcomp}{\voigtcomp}\right\}C\right]$ \\[20pt]
    \hline
    \hline
    \end{tabular}
\end{sidewaystable}

\clearpage
\bibliographystyle{yahapj}
\bibliography{references}

\end{document}